\newcommand{\g}{$\gamma$}

\newcommand{\cd}{C$_6$D$_6$}
\newcommand{\cds}{C$_6$D$_6$ }

\newcommand{\lacl}{LaCl$_{3}$(Ce)} 
\newcommand{\lacls}{LaCl$_{3}$(Ce) }
\newcommand{\fpis}{4$\pi$ i-TED }

\documentclass[epj]{svjour}
\usepackage{graphicx}
\usepackage{graphics}
\usepackage{epstopdf}
\usepackage{amssymb}   
\usepackage{url}
\usepackage{amsmath}

\begin{document}
\title{Imaging neutron capture cross sections: i-TED proof-of-concept and future prospects based on Machine-Learning techniques}

\author{
 V.~Babiano-Su{\'a}rez\inst{1} 
 \and J.~Lerendegui-Marco\footnote{\email{jorge.lerendegui@ific.uv.es}}\inst{1}
 \and J.~Balibrea-Correa\inst{1}
 \and L.~Caballero\inst{1} 
 \and D.~Calvo\inst{1}  
 \and I.~Ladarescu\inst{1} 
 \and C.~Domingo-Pardo\inst{1}
 \and F.~Calvi{\~n}o\inst{2}
 \and A.~Casanovas\inst{2}
 \and A.~Tarife\~no-Saldivia\inst{2}
% ...
}   

\institute{
Instituto de F{\'\i}sica Corpuscular, CSIC-University of Valencia, Valencia, Spain
\and Universitat Politecnica de Catalunya (UPC), Barcelona, Spain
% ...
}

\date{Received: \today / Revised version: \today}
%
\abstract{i-TED is an innovative detection system which exploits Compton imaging techniques to achieve a superior signal-to-background ratio in ($n,\gamma$) cross-section measurements using time-of-flight technique. This work presents the first experimental validation of the i-TED apparatus for high-resolution time-of-flight experiments and demonstrates for the first time the concept proposed for background rejection. To this aim both $^{197}$Au($n,\gamma$) and $^{56}$Fe($n, \gamma$) reactions were measured at CERN n\_TOF using an i-TED demonstrator based on only three position-sensitive detectors. Two \cds detectors were also used to benchmark the performance of i-TED. The i-TED prototype built for this study shows a factor of $\sim$3 higher detection sensitivity than state-of-the-art \cds detectors in the $\sim$10~keV neutron energy range of astrophysical interest. This paper explores also the perspectives of further enhancement in performance attainable with the final i-TED array consisting of twenty position-sensitive detectors and new analysis methodologies based on Machine-Learning techniques.
\PACS{
      {PACS-key}{discribing text of that key}   \and
      {PACS-key}{discribing text of that key}
     } 
} 
\maketitle

\begingroup
\let\clearpage\relax
\twocolumn
\endgroup


    
\section{Introduction}\label{sec:introduction}

 Neutron capture cross-section measurements are fundamental in the study of astrophysical phenomena, such as the slow neutron capture (s-) process of nucleosynthesis operating in red-giant stars~\cite{Kappeler11}. This mechanism is responsible for the formation of about half of the elements heavier than iron. One prominent method to measure neutron-capture cross sections over the full stellar range of interest is the time-of-flight (TOF) technique. Here, a sample of the isotope of interest is placed in a pulsed neutron beam and the prompt capture $\gamma$-rays originating from the sample are registered by means of radiation detectors. Low-efficiency \cds liquid scintillators are widely used to register the capture $\gamma$-rays~\cite{Macklin67,Plag03}. This type of detector, in conjunction with the pulse-height weighting technique (PHWT)~\cite{Macklin67} allows one to virtually mimic an ideal Total Energy Detector (TED)~\cite{Moxon63}. A TED is a detection system whose $\gamma$-ray detection efficiency becomes proportional to the registered $\gamma$-ray energy. Thanks to detailed Monte Carlo (MC) calculations~\cite{Tain02,Tain04,Borella07}, the working principle of the PHWT is so well under control nowadays that, generally, \cds detectors are also directly referred to as TEDs. Additionally, liquid scintillators are particularly convenient for neutron capture TOF experiments because of their of their fast time-response and low intrinsic sensitivity to\hfill \break
\hfill \break
\hfill \break
\hfill \break
\hfill \break
\hfill \break
\hfill \break
\hfill \break
\hfill \break
\hfill \break
\hfill \break
\hfill \break
\hfill \break
\hfill \break
\hfill \break
\hfill \break
\hfill \break
\hfill \break
\hfill \break
\hfill \break
\hfill \break
\hfill \break
\hfill \break
\hfill \break
\hfill \break
\hfill \break
\hfill \break
\hfill \break
\hfill \break
\hfill \break
\hfill \break
\hfill \break
\hfill \break
\hfill \break
\hfill \break scattered neutrons~\cite{Plag03,L6D6,Balibrea21}.
 
 However, an important limitation in many TOF experiments arises from neutrons that are scattered in the sample and get subsequently captured (prompt or after some thermalization) in the surroundings of the \cds detectors (see for instance Refs.~\cite{Domingo06,Tagliente13}). According to MC studies~\cite{Zugec14}, in the 1~keV to 100~keV neutron energy interval of relevance for astrophysics, this type of background may represent one of the dominant contributions in many neutron capture experiments.

In order to improve this situation a total-energy detector with $\gamma$-ray imaging capability, so-called i-TED, has been recently proposed~\cite{Domingo16}. i-TED exploits Compton imaging techniques with the aim of obtaining information about the incoming direction of the detected $\gamma$-rays. This additional information can help to reject events which do not arise directly from the capture sample under study, thereby enhancing the signal-to-background ratio (SBR). This novel detection system is under development at IFIC, a first demonstrator has been assembled and its main components and techniques have been characterized and optimized in the recent years~\cite{Olleros18,Babiano19,Babiano20,Balibrea20}.

This article discusses the status and future perspectives for background rejection with i-TED and it consists of two parts. In the first part (Sec.~\ref{sec:Proof}) we present a proof-of-concept (PoC) experiment carried out with an i-TED demonstrator. The latter is intended to technically validate its overall performance for TOF experiments and to demonstrate the background rejection concept. The PoC experiment was carried out at CERN n\_TOF, which is introduced in Sec.~\ref{sec:nTOF}. The i-TED demonstrator assembled for the present study consisted of 3 position-sensitive detectors (PSDs) as described in Sec.~\ref{sec:iTED}. Its energy- and spatial-sensitivities are discussed in Sec.~\ref{sec:calibrations}. The experimental validation of the i-TED prototype for TOF experiments is reported in Sec.~\ref{sec:TOFiTED}. Finally, Sec.~\ref{sec:iTEDsensitivity} describes the background rejection results and the limitations in detection efficiency when the so-called analytical approach~\cite{Domingo16} is applied. Empowered by these results, we have investigated alternative analysis methodologies, which allow one to preserve a better detection efficiency. These are reported in the second part of this article, which starts in Sec.~\ref{sec:Prospects}. First, we discuss the prospects for the final detector~\cite{Domingo16} consisting of 20 PSDs arranged in four Compton modules. The perspectives for background rejection using Compton imaging techniques with the final \fpis detector are presented in Sec.~\ref{sec:ImagingiTED5}. Finally, Sec.~\ref{sec:MLBackground} explores the new analysis methodology for the background rejection with i-TED, which is based on Machine-Learning techniques and allows one to remarkably improve its performance, when compared to the analytical method previously used. A summary and outlook of our results is provided in Sec.~\ref{sec:summary}.

\section{i-TED Proof-of-concept}\label{sec:Proof}

The objective of the PoC experiments was twofold. On one hand, the aim was to investigate the suitability of i-TED for neutron-capture TOF experiments, whereby a sufficiently fast detector response is mandatory to preserve the TOF or neutron-energy resolution. On the other hand, the goal was to demonstrate the applicability of the proposed $\gamma$-ray imaging technique to suppress spatially localized $\gamma$-ray background sources.

Hence, two different neutron capture experiments were carried out at the CERN n\_TOF facility~\cite{Gunsing17}. This installation and the conventional experimental set-up of two \cds detectors~\cite{L6D6} are described in Sec.~\ref{sec:nTOF}. The \cds detectors represent the state-of-the-art in this field and thus they are systematically used as reference for comparison purposes along this work. The i-TED demonstrator is described in Sec.\ref{sec:iTED}. The reconstruction of the \g-ray energy depositions and interaction points in i-TED were based on our previous works~\cite{Olleros18,Babiano19,Babiano20} and for more details the reader is referred to those articles and references therein. Sec.~\ref{sec:calibrations} shows the performance of the imager in terms of energy and Compton imaging capability by the time of the experiment. The first PoC experiment is described in Sec.~\ref{sec:TOFiTED}. Here, a simultaneous measurement of the $^{197}$Au($n,\gamma$) reaction both with \cds detectors and with i-TED was performed in order to demonstrate the suitability of i-TED for the neutron-TOF technique in a high-resolution facility. The second PoC experiment is described in Sec.~\ref{sec:iTEDsensitivity} and consisted of a simultaneous measurement of the $^{56}$Fe($n,\gamma$) reaction with both \cds detectors and i-TED. The objective here was to demonstrate the background rejection capability of i-TED and to quantify the attainable enhancement in terms of signal-to-background ratio (SBR). This reaction was chosen because it shows an isolated resonance at $E_\circ=1.15$~keV, which is well suited to evaluate SBRs in the several $\sim$10~keV of neutron energy, which is the range of interest for astrophysics.

\subsection{The n\_TOF facility and experimental set-up}\label{sec:nTOF}    

The CERN proton synchrotron (PS) delivers a pulsed beam of protons (7$\times$10$^{12}$ protons/pulse, 7~ns \textsc{rms}) with a typical duty cycle of 0.25~Hz and an energy of 20 GeV. This beam impinges onto a lead target generating about 300 neutrons per incident proton by means of spallation reactions. Most unwanted charged particles generated from the spallation reactions are removed from the beam by using magnets and collimators, which do not prevent a significant portion of \g-rays to remain in it. After crossing a water moderator which surrounds the spallation source, a white neutron spectrum becomes available for experiments at two experimental areas EAR1 and EAR2, which are located at 185 m and 20 m distance from the source. In both cases, the energy of the incoming neutrons is determined by measuring their TOF. The different distances of the flight paths yield a better energy resolution at EAR1, and a higher neutron flux at EAR2. A prompt and intense flash of $\gamma$-rays arrives at both experimental areas few nanoseconds after being created by the proton beam impact. This flare is employed for the conventional C$_6$D$_6$ detectors to time-stamp the beginning (t$_{\circ}$) of the neutron bunch.

\begin{figure}[!htb]
   \begin{center}     
   \includegraphics[width=\columnwidth]{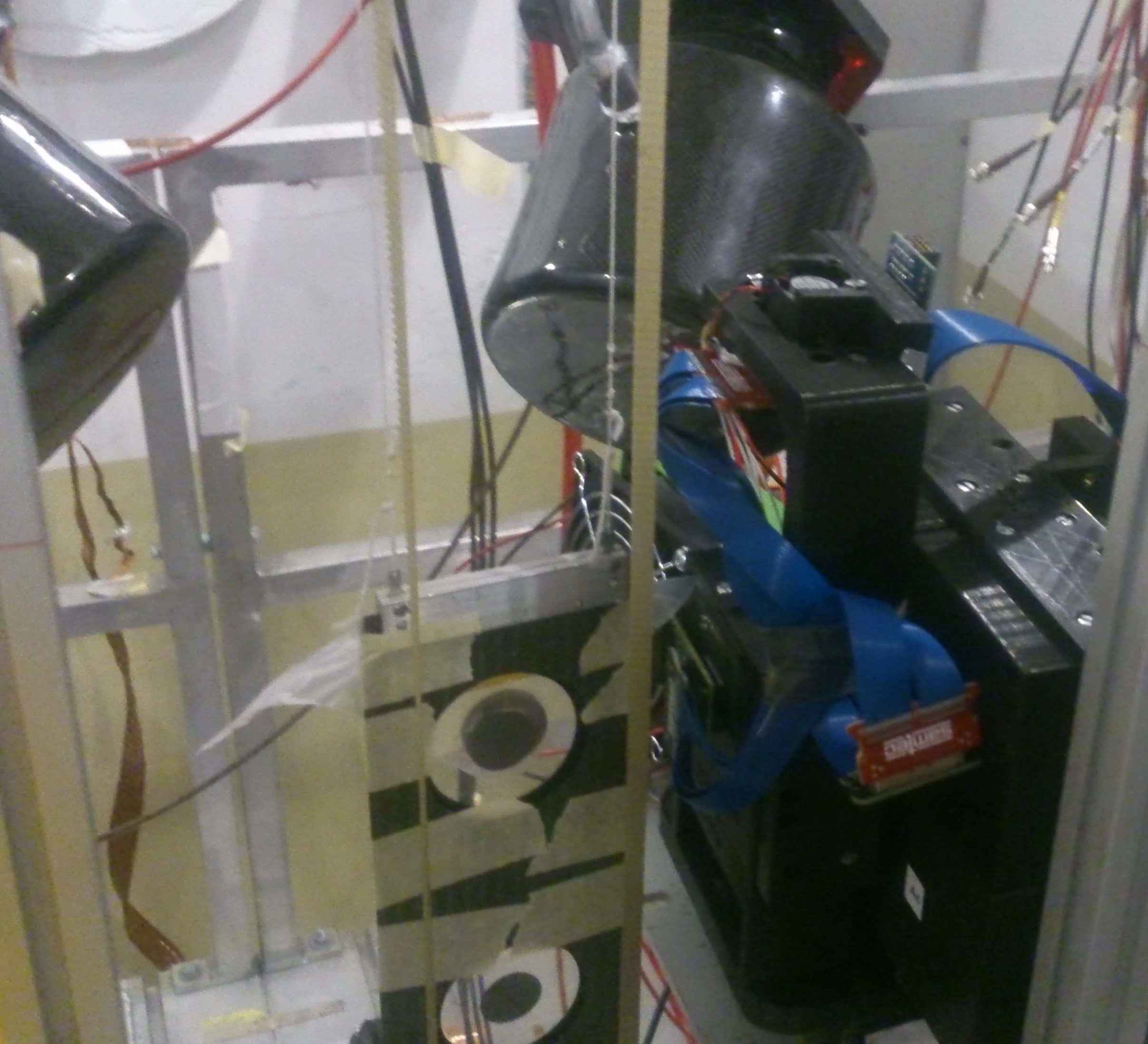}
   \end{center}
\caption{\label{fig:ExpSetup} Photograph of the experimental setup. Two C$_6$D$_6$ (partially visible in the top region) were combined with the i-TED demonstrator (right side of the image). Several capture samples were placed in the sample exchanger. See also Fig.~\ref{fig:iTED} for a clearer view of the \g-ray imager.}
\end{figure}

The PoC experiments were carried out at the EAR1 experimental area of n\_TOF. A picture of the measuring setup is shown in Fig.~\ref{fig:ExpSetup}. The i-TED demonstrator was placed at an angle of about 90$^{\circ}$ with respect to the beam and at 63~mm distance from the center of the capture sample under study. The distance between the two position-sensitive detection planes of i-TED, so called Scatter and Absorber (see Sec.~\ref{sec:iTED}) was kept fixed to 10~mm, which represented a reasonable trade-off between angular resolution and efficiency~\cite{Babiano20}. In the neighbourhood of i-TED two C-fibre C$_6$D$_6$ detectors were set up. The latter detectors are used as reference along the whole work. 

The two C$_6$D$_6$ detectors were oriented at an angle of 60$^{\circ}$ with the vertical, with their faces pointing to the capture sample placed at a distance of 10 cm. The two photomultiplier tube (PMT) readout channels are connected to the n\_TOF digital acquisition system~\cite{Abbondanno05}. The acquisition system of i-TED is described in Sec.~\ref{sec:iTED}. The i-TED data was stored in a PC located in the experimental hall and remotely controlled via Ethernet connection.

Samples of gold and iron were placed in the neutron beam for these PoC measurements. Gold is commonly employed as reference for the normalization of the capture yield via the Saturated Resonance method~\cite{Macklin67}. However, in this case gold was only measured to validate the TOF calibration and response of both detection systems. On the other hand, the iron sample (enriched 99\% in $^{56}$Fe) was measured because of its high scattering-to-capture ratio and its isolated resonance at 1.15~keV. This resonance, in conjunction with the background "plateau" at $\sim$10~keV mostly induced by scattered neutrons was used to systematically evaluate the SBR for both detection systems and under different analysis conditions. Finally, to monitor the detector gain along the experiment a $^{152}$Eu radioactive source with 14~kBq of activity was placed at about 10~cm distance from i-TED. 

\subsection{The i-TED demonstrator}\label{sec:iTED}    
\begin{figure}[!htbp]
   \begin{center}    
   \includegraphics[width=5.75cm]{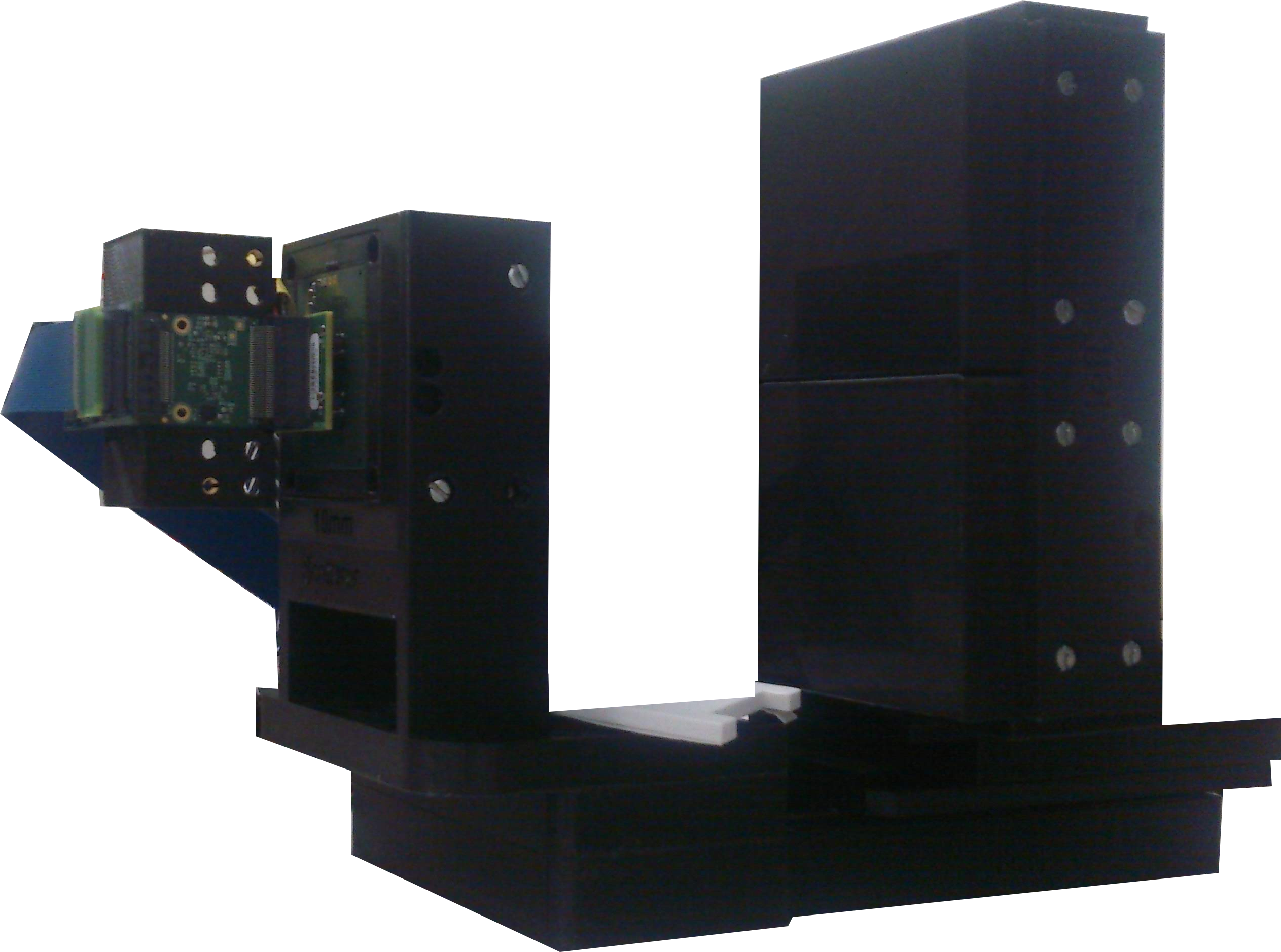}
   \end{center}
\caption{\label{fig:iTED} The i-TED5.3 demonstrator equipped with one scatter and two absorber detectors in movable and parallel detection planes.}
\end{figure}

The imager used in this work consists of a Compton camera that utilizes three PSDs distributed in two parallel detection planes, Scatter (S) and Absorber (A), as shown in Fig.~\ref{fig:iTED}. Each PSD contains a LaCl$_{3}$(Ce) monolithic crystal with a square-cuboid shape and a base surface of 50$\times$50~mm$^2$. The \lacls is hygroscopic and thus it is encapsulated in an aluminum housing. The crystal base is coupled to a 2~mm thick quartz window, which is optically coupled to a silicon photomultiplier (SiPM) from SensL (ArayJ-60035-64P-PCB). The photosensor features 8$\times$8 pixels over a surface of 50$\times$50~mm$^2$. A 10 mm thick crystal is used for the PSD in the S-plane. Two 25 mm thick crystals are utilized for the two PSDs placed in a vertical stack for the A-plane (see Fig.~\ref{fig:iTED}). In total, 192 SiPM channels are biased and readout by means of front-end and processing PETsys TOFPET2 ASIC electronics~\cite{PETsys16}. In order to minimize gain shifts due to changes in the temperature of the experimental hall, every ASIC is thermally coupled to a refrigeration system composed by Peltier cells (see Ref.~\cite{Babiano20} for further details). 
On the other hand, since i-TED is intended for TOF measurements, the PETsys readout electronics had to be supplemented with an external module especially designed to provide an accurate time-stamp for the time reference of each neutron bunch. The latter device was based on the high-speed differential-line driver SN65LVDS9638 from Texas Instruments~\cite{lvds}. This chip features a propagation delay of only 1.7~ns and delivers a Low-Voltage Differential Signal (LVDS) with output rise- and fall-times of only 500~ps. Using this trigger-module, every TTL-trigger signal from the CERN PS was converted into a LVDS-pulse, which was fed into the PETsys communication mezzanine via two available ports. The time of arrival of the latter trigger signal was digitized with a 20~ps time resolution using one of the 192 ASIC channels. In practice we experienced jitter issues with this external trigger module, which apparently were due to an insufficient electromagnetic compatibility in the harsh conditions of the n\_TOF environment. These difficulties could be circumvented to a large extent by means of a software-correction algorithm applied to the raw data. However, this correction is rather complex and not fully accurate and thus, for future experiments a hardware solution will be pursued.

\subsection{Energy calibration and $\gamma$-ray imaging validation}\label{sec:calibrations}
\begin{figure}[!htbp]
   \begin{center}     
   \includegraphics[width=0.9\columnwidth]{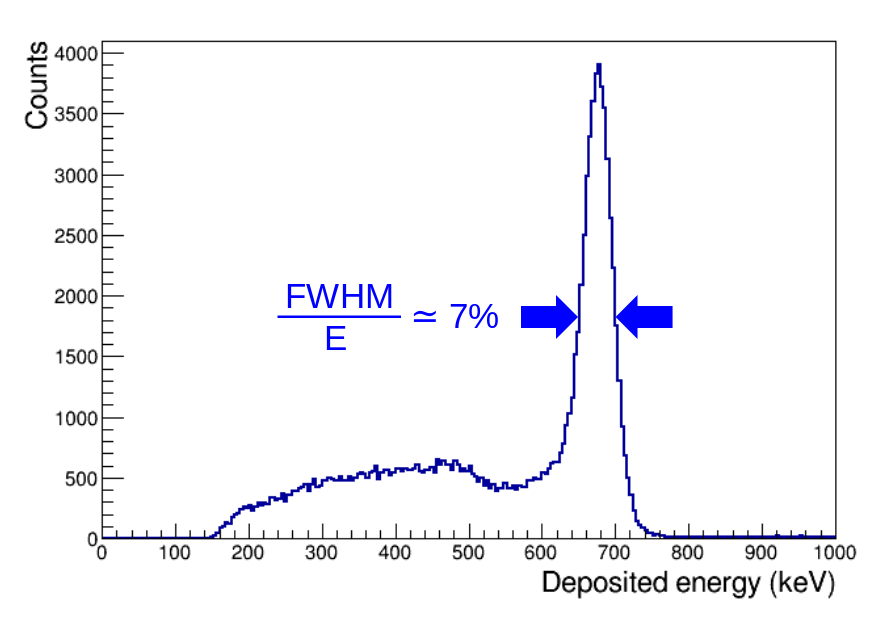}
   \end{center}
\caption{\label{fig:EdepCs} Add-back spectrum for a $^{137}$Cs source with the S- and A-layers in time-coincidence.}
\end{figure}

Deposited-energy values and 3D-localization of the $\gamma$-ray hit position in each \lacls crystal are necessary to apply the Compton scattering law on an event-by-event basis and obtain information on the incoming $\gamma$-ray direction~\cite{Babiano20}.
An energy calibration of every i-TED PSD in the energy range between 122~keV and 1408~keV was performed by placing a $^{152}$Eu radioactive source near both detection planes. Additionally, the intrinsic $\alpha$-activity of the LaCl$_{3}$(Ce) crystals~\cite{Hartwell05} was used to extend the calibration range up to 2.8~MeV. Finally, capture \g-ray spectra of the gold and iron samples measured at CERN n\_TOF with i-TED were used, in combination with a MC-simulation of the capture-cascades, to extend the calibration range up to 7.6~MeV. A combination of two second-degree polynomials were used as  energy-calibration function for each PSD in the full energy range of the neutron-capture experiments. Energy resolutions between 7\% and 5\% \textsc{fwhm} at 662 keV have been found for the crystals placed in the S- and in the A-layers, respectively. Operating both layers in time-coincidence mode, a resolution of $\sim$7\% \textsc{fwhm} was found at 662~keV for the add-back spectrum of a $^{137}$Cs source (see Fig.~\ref{fig:EdepCs}).
The 2D-coordinates of the $\gamma$-ray interaction in the transverse $xy$-plane of each PSD in i-TED is accomplished by fitting, on an event-by-event basis, the analytical expression of Li et al.~\cite{Li10} to the SiPM charge response of each PSD, in a similar fashion as reported in Ref.~\cite{Babiano19}.  After a careful characterization of all detectors, spatial resolutions of around 1 mm \textsc{fwhm} were found for the 10~mm and 25~mm thick crystals at 511~keV. The third space coordinate $z$ or depth-of-interaction in each crystal is obtained from an empirical parameterization of the SiPM response at half-height of the maximum\cite{Babiano20}. In the latter case, typical uncertainties between 4~mm and 5~mm are obtained. The field of view in which the PSD has a linear response was determined to be of around 20~cm$^2$, which corresponds to about 80\% of the crystal base surface~\cite{Babiano19}.  

\begin{figure}[!htbp]
   \begin{center}     
   \includegraphics[width=\columnwidth]{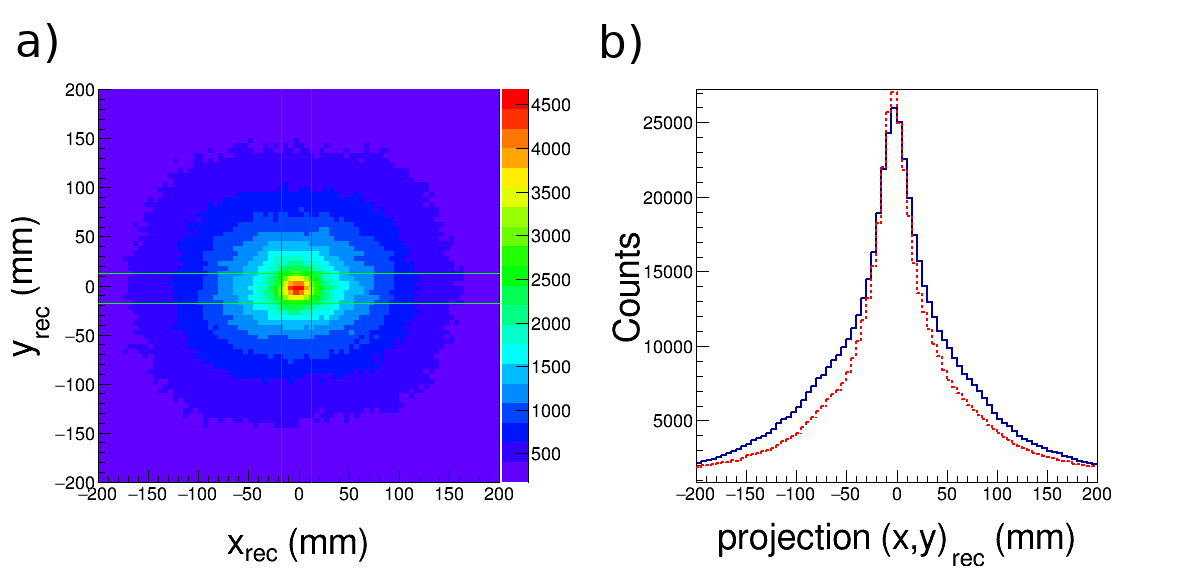}
   \includegraphics[width=\columnwidth]{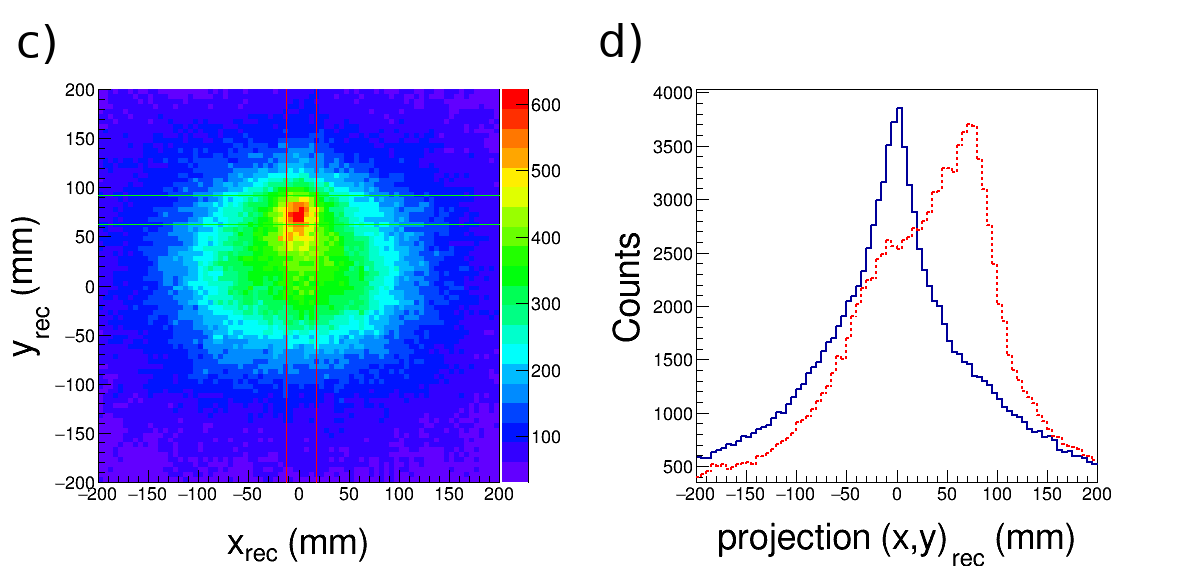}
   \end{center}
\caption{\label{fig:CsCompton} Reconstructed Compton images of a $^{137}$Cs source placed on capture sample position (a), and 10~cm up (c). The lines indicate the projection cuts. b) and d) show the projections of these cuts on both x- and y-axis.}
\end{figure}

A simple back-projection Compton image reconstruction algorithm was implemented~\cite{Babiano20} in order to validate the overall \g-ray imaging performance of i-TED, which relies on accurate energy- and spatial-calibrations. Thus, a point-like $^{137}$Cs source with an activity of 325~kBq was attached to one of the capture samples at a distance of 63~mm in front of the S-detector of i-TED (see also Fig.~\ref{fig:ExpSetup}). In a second measurement, the $^{137}$Cs-source was lifted 10~cm up in the vertical direction. The resulting Compton images are shown in Fig.~\ref{fig:CsCompton}, thereby illustrating the angular sensitivity of the i-TED imager and the proper implementation of calibrations and spatial reference systems.

\subsection{TOF performance of i-TED}\label{sec:TOFiTED}
In order to benchmark the TOF performance of i-TED with respect to \cds detectors we measured a sample of $^{197}$Au with a diameter of 20~mm, a thickness of 0.125~mm and a mass of 645~mg. Gold has the advantage that its resolved-resonance region is well known~\cite{Massimi10,Carlson18} and the narrow resonances can be used to calibrate the TOF-$E_n$ relation and to verify the TOF-resolution.

\begin{figure*}[!hbtp]
   \begin{center}     
\includegraphics[width=\textwidth]{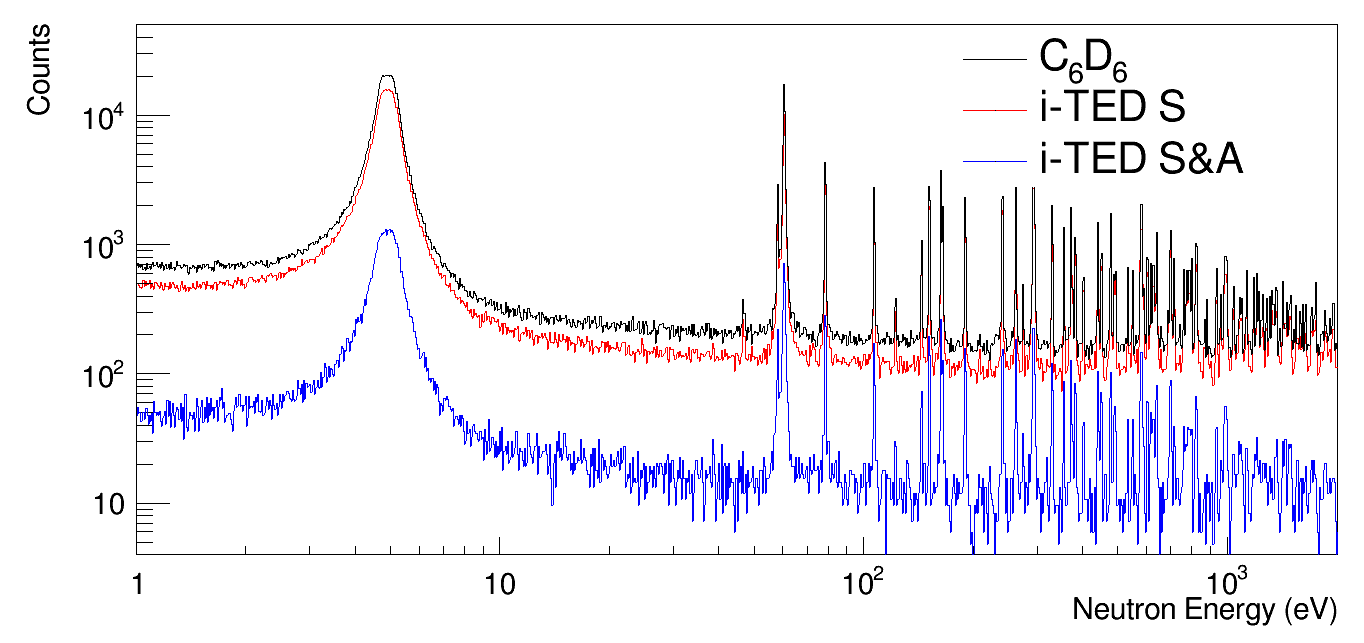} 
    \end{center}
\caption{\label{fig:197AuEnSpectra} Neutron-energy spectra (300 bins per decade) measured for the $^{197}$Au capture sample with two \cds detectors and with i-TED. See text for details.}
\end{figure*}

The TOF is measured as the time-interval between the origin of the neutron bunch, $t_{\circ}$, and the time of detection of any of the capture \g-rays arising from the sample under study, $\Delta t = t_{\gamma}-t_{\circ}$. With \cds detectors the  $t_{\circ}$-value can be reliably determined from the signature of the so-called \g-flash, which are the prompt \g-rays produced in the spallation reactions~\cite{Abbondanno05}. In the latter case, $t_{\circ} = t_{flash} - t_{corr}$, where $t_{corr}$ takes into account the time needed for the light to travel along the flight-path of 185~m. On the other hand, the acquisition system of i-TED does not allow to digitize the whole movie of pulses and thus an alternative methodology had to be implemented. The latter consisted of time-stamping in the i-TED acquisition system the rising-edge of the external PS-trigger TTL-signal, $t_\textsc{ps}$, as described in Sec.~\ref{sec:iTED}. Thus, for i-TED the TOF of each registered \g-ray was determined as $\Delta t = t_{\gamma}-t_{\textsc{ps}} + t_{\text{offset}}$. The offset-correction $t_\text{offset}$ is due to the time-difference between the PS-trigger signal and the actual time $t_{\circ}$ of the proton-beam on the target. The value of $t_{\text{offset}}$ was determined empirically from the wealth of resonances in the $^{197}$Au+n spectrum measured with i-TED during these experiments. After these considerations, the TOF is converted into neutron energy by means of the relationship
\begin{equation}
    E_n = \alpha^2 \frac{L^2}{\Delta t^2}
\end{equation}
where $L$=185~m is the flight-path and $\alpha$=72.29~$\sqrt{eV}\mu s/m$ with $\Delta t$ expressed in $\mu$s. 

Fig.~\ref{fig:197AuEnSpectra} shows the neutron-energy spectra obtained with both i-TED and one C$_6$D$_6$ detector for the $^{197}$Au(n,$\gamma$) reaction. All spectra correspond to the same measuring time and thus, differences in counting statistics are only due to the different intrinsic and geometric efficiencies of each detector. This is an important aspect, for it shows that the S-detector of i-TED placed at 63~mm distance from the sample has an efficiency which is about 75\% of the one obtained with a \cds detector placed at 10~cm. Thus, if one assumes similar sample-detector distances in future experiments, the four S-detectors of the \fpis array will provide an efficiency comparable to that of the conventional \cd-array of four detectors. On the other hand, the efficiency of the i-TED prototype in coincidence mode is only 6.3\% of the one shown by the one \cds detector. Considering that the demonstrator had only two A-detectors instead of the four intended for the final \fpis detector, according to our MC calculations for the latter (Sec.~\ref{sec:MCiTED5}) one can expect an efficiency between 10\% and 14\% of that obtained with four \cds detectors. The precise value depends on the separation between the S- and A-layers. The additional interplay between detection efficiency and background rejection will be discussed in detail in Sec.~\ref{sec:iTEDsensitivity} and in the second part of this paper.

\begin{figure}[!htbp]
\includegraphics[width=0.9\columnwidth]{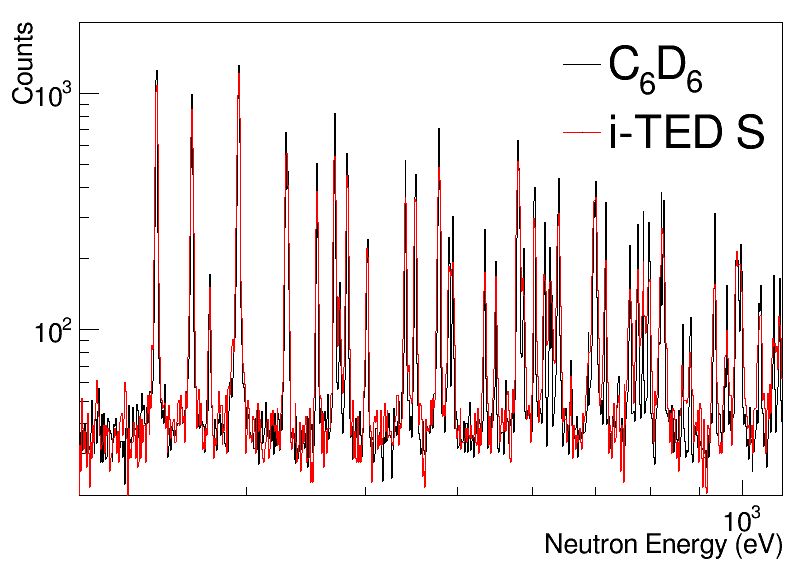} 
\caption{\label{fig:Auzoom}Expansion of the neutron-energy spectra of \cds detectors and i-TED in the 200~eV-to-1~keV neutron energy range (1000 bins per decade).}
\end{figure}

In terms of TOF resolution, a good agreement was found for both detection systems. This result was not evident beforehand, given the large differences in terms of scintillation materials (\lacls and \cd), readout photo-sensors (SiPM and PMT) and processing electronics (ASICs and digitizer modules). Fig.~\ref{fig:Auzoom} shows an expansion of the energy range between 200~eV and 1~keV using 1000 bins per decade. A more detailed assessment of the possible detector contribution to the TOF-resolution would require much higher counting statistics and additional measurements of very narrow resonances over the entire neutron-energy range, preferably employing samples of $^{32}$S and $^{238}$U. This aspect will be investigated in more detail in future measurements with the entire \fpis detection system, where the efficiency will be sufficient to perform such studies in a reasonable time span.
At this stage, one can conclude that the fast time-response of the PSDs embedded in i-TED and its readout electronics is satisfactory for performing high-resolution TOF experiments with pulsed neutron beams.



\subsection{Background rejection with i-TED}\label{sec:iTEDsensitivity}
The second PoC experiment was the measurement of the $^{56}$Fe($n,\gamma$) reaction at n\_TOF EAR1 using the same experimental set-up. This isotope was chosen because it has a narrow and isolated resonance at $E_{\circ}$=1.15~keV and elastic scattering overcomes capture in the 1~keV-to-10~keV energy range by about four orders of magnitude. The large fraction of neutrons scattered in the iron sample are expected to undergo further scattering in the surroundings of the set-up until they are eventually captured, mainly in the concrete walls of the experimental hall~\cite{Zugec14}, emitting further background radiation.  Thus, this experimental situation is particularly well suited to explore the capability of i-TED to reject background \g-ray events that do not emerge from the capture sample.

\begin{figure*}[!htbp]
   \begin{center} 
   \includegraphics[width=\textwidth]{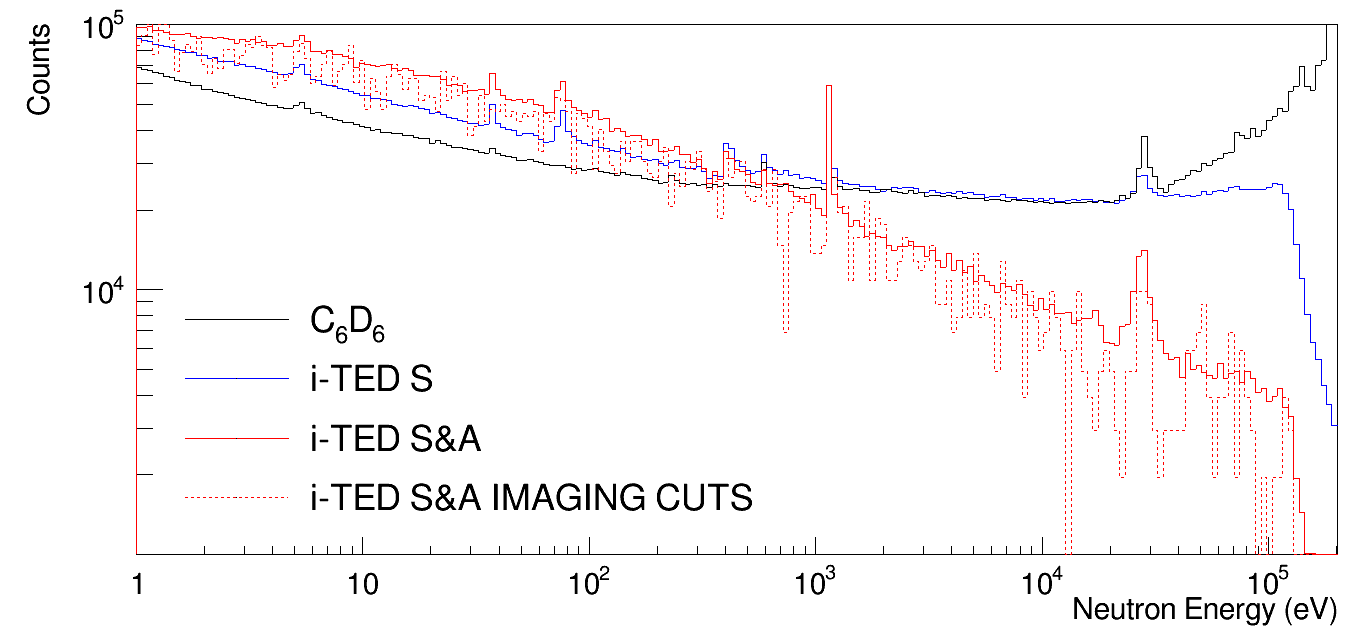}
    \end{center}
\caption{\label{fig:EnFeiTEDvsC6D6imaging} Neutron energy spectra (40 bins per decade) measured with the $^{56}$Fe sample using the i-TED S-detector in singles-mode (blue line) and i-TED with S- and A-detectors in time-coincidence mode (solid red line). The spectrum measured with the C$_6$D$_6$ detectors is shown in black. The dashed-red spectrum shows the best result obtained with \g-ray imaging cuts. See text for details.}
\end{figure*}

The iron sample was isotopically enriched in $^{56}$Fe (99.93\%), it had a diameter of 20~mm and a mass of 2.1035~g. Fig.~\ref{fig:EnFeiTEDvsC6D6imaging} shows the neutron energy spectra of counting statistics obtained with \cds detectors and with i-TED using only 40 bins per decade. In this case a rather coarse binning had to be used due to the low cross section and the limited counting statistics. For the sake of comparison all spectra have been normalized to the peak of the 1.15~keV resonance in $^{56}$Fe+n. 

Let us discuss first the singles-response measured with the S-detector of i-TED shown with a solid-blue line in Fig.~\ref{fig:EnFeiTEDvsC6D6imaging}. In the neutron energy range from 1~eV up to 1~keV the background level of the S-detector is a factor of $\sim$2 higher than that of the \cds detectors. This is mainly due to the intrinsic $\alpha$-activity of the \lacls crystals, which dominates the counting rate in the low neutron-energy range. In addition, a few neutron-capture resonances can be observed that do not belong to $^{56}$Fe+n. The peak at 5.2~eV corresponds to a small silver impurity ($\leq$100~ppm) in the iron sample and thus it appears with a similar strength in all spectra. Other prominent contaminants are found mainly in the i-TED data at 72.3~eV, 398~eV and 617~eV. These peaks correspond to the largest  neutron capture resonances in $^{139}$La and $^{35}$Cl, and are the signature of the intrinsic neutron sensitivity of the \lacls crystals in this neutron-energy range. Although inconvenient, this type of background cannot be considered yet an important limitation. On one side, for the final system we plan to use a neutron absorber made from polyethylene enriched in $^{6}$Li, which is expected to reduce significantly this effect, while becoming essentially transparent to \g-rays~\cite{Domingo16}. Similar techniques are conventionally used for the same purpose in experiments using the Total-Absorption Calorimeter (TAC)~\cite{Guerrero2008}. On the other hand, the higher neutron-energy range is more relevant for $s$-process studies dealing with hydrogen burning ($kT$=8keV) and He-shell flashes ($kT$=23keV) in AGB-stars, and during core He-burning ($kT$=26~keV) and shell C-burning ($kT$=90~keV) in massive stars. Experimentally, it is indeed in the 1~keV to 100~keV neutron energy range where there is a need for enhancing the SBR with respect to \cds detectors, owing to the 1/$v$-dependence of the cross sections and the increasing contribution of the scattered-neutrons induced \g-ray background~\cite{Zugec14}.
At the 1.15 keV resonance, the SBR is comparable for both C$_6$D$_6$-detectors and the S-detector of i-TED. This indicates a similar performance between them at this neutron energy. The background level remains more or less constant and similar for both \cds detectors and the S-detector of i-TED up to the second strong resonance of $^{56}$Fe+n at 30~keV. 
Let us compare now the i-TED spectrum in time-coincidence between the S- and any of the A-detectors (S\&A), with respect to \cds detectors. The first remarkable aspect is that below $E_n\sim$1~keV the background of i-TED increases with respect to the \cds detectors and with respect to the S-detector. To a large extent this is due to the bottom \lacls A-crystal, which had an anomalously large contribution of $\alpha$-emitting isotopes, yielding an $\alpha$-counting rate of $\sim$1~kHz in that particular crystal. This activity is about one order of magnitude higher than average values. However, by the time of the experiment there was no other crystal available for replacement and efficiency was a major concern.

Beyond neutron energies of $E_n\sim$1~keV one can observe a decreasing trend in the background of i-TED when compared to the \cds detectors and to the S-detector. This effect was expected from the MC-simulations (see Sec.~\ref{sec:iTED4piImaging}) and it is ascribed to the configuration of i-TED and the nature of the background in this energy range~\cite{Zugec14}. Firstly, neutron-induced background \g-rays coming from the walls into the detector have a rather soft energy spectrum compared to capture \g-rays and thus, the thick A layer of i-TED acts as a veto for a large portion of these events. Moreover, in-beam \g-rays scattered in the iron sample, featuring a soft spectrum dominated by 478~keV \g-rays ~\cite{LoMeo15}, are also significantly shielded by the S crystal.
In contrast to i-TED, the \cds detectors register in equal measure all the \g-rays which reach their sensitive detection volume, regardless of their origin and hardness. As a consequence of the different sensitivity to the background components, the level of the background in i-TED operating in time-coincidence mode is reduced by a factor of 2.5 at 10~keV with respect to \cds detectors. Consequently, the SBR of the $^{56}$Fe(n,$\gamma$) resonances have been noticeably improved from a factor of 2 in both resonances for the C$_6$D$_6$, to a factor of 3 and 2.6 for i-TED in both resonances, respectively. 

Finally, in order to explore the reduction in background by means of the $\gamma$-ray imaging capability of i-TED a discrimination analysis was made on an event-by-event basis. To apply the Compton technique only events in time coincidence between the S- and A-layers were considered (solid red curve in Fig.~\ref{fig:EnFeiTEDvsC6D6imaging}). The main drawback of the time-coincidence mode is the drastic reduction of the detection efficiency. As a consequence of the small number of A-detectors used in the present prototype, only 8\% of the events detected in the S-detector make a further interaction in the A layer. In future measurements, the use of four PSDs in the A- plane will help to enhance the efficiency in coincidence mode significantly (see Ref.~\cite{Babiano20}).

\begin{figure}[!htbp]
   \begin{center} 
   \includegraphics[width=\columnwidth]{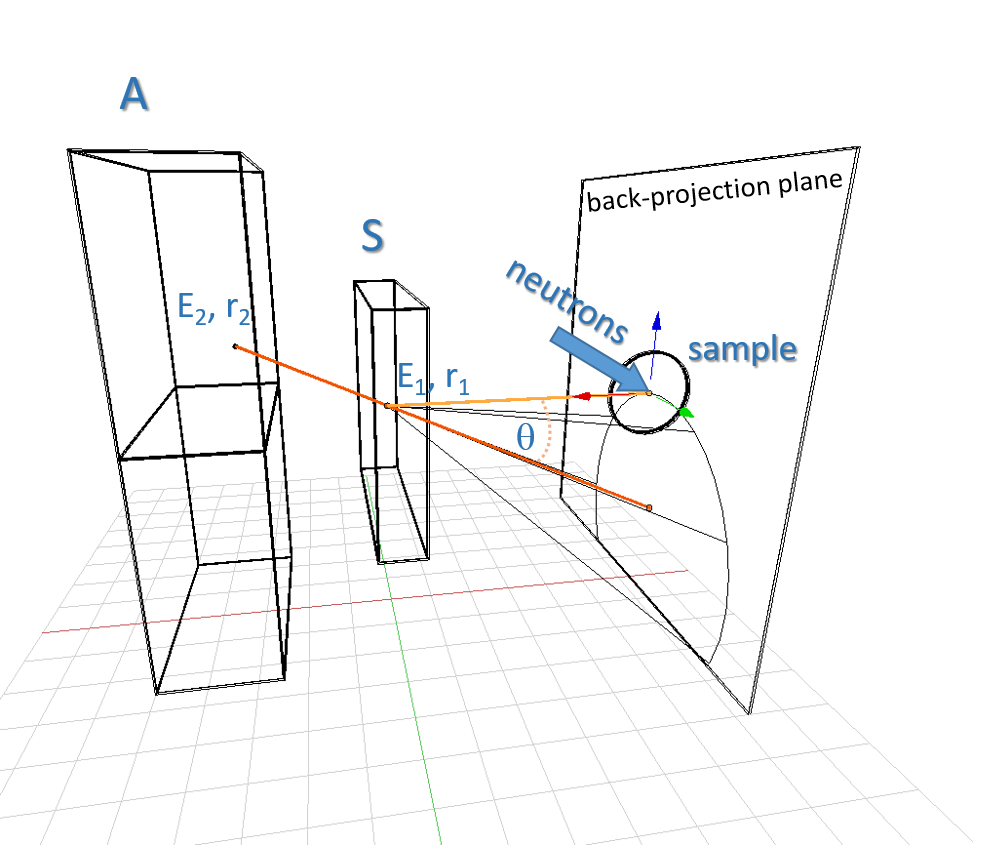}
      \end{center}
\caption{\label{fig:iTEDconcept} Schematic representation of a capture event validated with i-TED by means of the Compton technique. Measurable quantities ($E_1$, $E_2$, $r_1$ and $r_2$) correspond to the interaction of the capture \g-ray in both detection layers.}
\end{figure}

Deposited energies $E_1$ and $E_2$ respectively in the S- and A-layers, and 3D-localization of the \g-ray hits, $\vec{r_1}$ and $\vec{r_2}$ in both layers were reconstructed as described in Sec.~\ref{sec:calibrations} for each coincidence event. From these quantities, one can trace a cone, whose central axis corresponds to the straight line defined by the \g-ray interaction position in the two layers and its aperture $\theta$ is obtained from the measured energies using the Compton scattering formula (see Fig.~\ref{fig:iTEDconcept}).
The aim of this procedure is to filter out events whose reconstructed Compton cone does not intersect with the sample position within a certain uncertainty range. To perform this spatial discrimination, the quantity $\lambda$, defined in Eq.~(\ref{eq:Lambda}), can be employed~\cite{Domingo16}. 
\begin{equation}
  \begin{split}\label{eq:Lambda}
\lambda = abs \Bigg[&(n_x a_x + n_y a_y + n_z a_z)^2  \\
& - \left(1 + \frac{511}{E_1 + E_2} - \frac{511}{E_2}\right)^2 (a_x^2 + a_y^2 + a_z^2)\Bigg].
 \end{split}
\end{equation}
In this equation, $n_x,n_y,n_z$ are the components of the unit vector along the cone axis defined by $\vec{r_1} - \vec{r_2}$, where $\vec{r_1} = (a_x,a_y,a_z)$. When $\vec{r_2}$ is expressed in mm, $\lambda$ has units of mm$^2$. For convenience, the origin of the Cartesian coordinates system is defined at the center of the sample position.
The $\lambda$-parameter represents the solution for the quadratic describing the intersection of the Compton cone with a plane at the sample center and perpendicular to the circular sample surface.  For a point-like \g-ray source located at the origin the cone coincides with the sample position for $\lambda = 0$. Because of the finite sample size and taking into account also instrumental uncertainties both on position and energy, in actuality one may expect good or true capture events for values of $\lambda$ in a region around $\lambda \gtrsim 0$. As described below, in this work we determine empirically the best choice of $\lambda$ for an optimal SBR.

\begin{figure}[!htbp]
   \begin{center} 
   \includegraphics[width=\columnwidth]{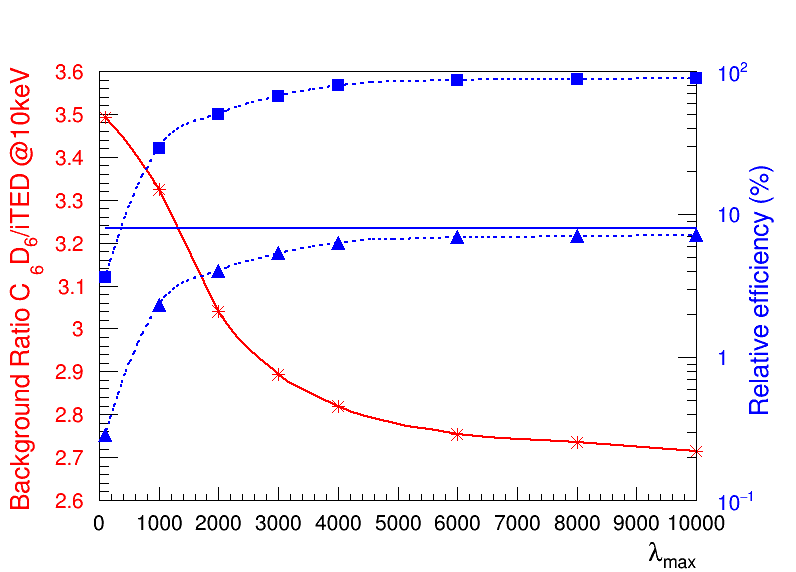}
    \end{center}
\caption{\label{fig:EfficiencySBratioFeiTED} Background ratio of \cds over i-TED at 10 keV of neutron energy in the $^{56}$Fe($n,\gamma$) spectrum (left axis and solid-red line) as a function of the $\lambda$ selection indicated by the maximum value of $\lambda$ in the horizontal axis ($\lambda_{max}$). The relative detection efficiency as a function of $\lambda_{max}$ is indicated by dashed-blue lines (right-hand axis). Solid squares represent the relative loss of efficiency as a function of $\lambda_{max}$. Solid triangles represent the relative efficiency with respect to the singles efficiency of the i-TED S-detector (8\%), which is indicated by the horizontal solid line.}
\end{figure}

On top of the time-coincidence technique, which leads to the aforementioned 8\% coincidence efficiency relative to the S-singles efficiency, the selection in the $\lambda$-distribution leads to an additional decrease in efficiency. This is shown in Fig.~\ref{fig:EfficiencySBratioFeiTED}, which shows the variation of relative efficiency as a function of the selection on $\lambda$, from $\lambda=0$ up to a maximum value of $\lambda_{max}$. The dashed-blue line with square symbols represents the fraction of counts in the $^{56}$Fe(n,$\gamma$) neutron-energy spectrum as a function of the selection on the $\lambda$ parameter, taking as reference (100\%) the situation where no cut is applied. The latter corresponds to the solid-red line spectrum in Fig.~\ref{fig:EnFeiTEDvsC6D6imaging}. The dashed-blue line with triangles shows the same result taking as reference the number of single events in the S-detector of i-TED. Since the efficiency of the S-detector is comparable to that of one \cds detector (see Sec.~\ref{sec:TOFiTED}), the latter line is a good representation of the efficiency for the i-TED prototype used in this work, compared with state-of-the-art systems (\cds detectors) as a function of the imaging cut applied to analyze the data. The maximum value (solid blue line) represents the fraction of the S-detector total counts which are selected in time-coincidence mode (the aforementioned $\sim$8\%). Finally, the SBR obtained for the $^{56}$Fe capture spectrum, as defined before using the background levels at 10~keV as reference, is displayed with the red line in Fig.~\ref{fig:EnFeiTEDvsC6D6imaging}. 
As expected, both efficiency and SBR curves follow an opposite trend. A stringent angular cut around the sample position corresponds to small $\lambda$-values, which provide an enhancement in SBR and a consequent decrease in counting statistics. The detection efficiency with cuts in imaging rapidly decreases for lower values of $\lambda$, becoming lower than 1\% for a lambda cut below 500~mm$^2$. A background ratio \cds over i-TED of $\sim$3 corresponds to a relative efficiency of about 4\% and a $\lambda$-cut of 2000 units. A more restrictive cut of $\lambda \leq$1000 increases the background ratio to about 3.3 and yields a relative efficiency of about 2\%. Last, the maximum background ratio of $\sim$3.5 is achieved with $\lambda \leq$100 at the cost of reducing the relative efficiency to just 0.2\%.

The resulting neutron energy spectrum of $^{56}$Fe($n,\gamma$) for an illustrative imaging selection of $\lambda \leq$500 is shown (red-dashed line) in Fig.~\ref{fig:EnFeiTEDvsC6D6imaging}. For the sake of comparison, this spectrum has also been normalized to the top of the 1.15~keV resonance. An enhancement in SBR becomes apparent in the 1~keV-30~keV neutron energy range, although the total counting statistics after the selection on $\lambda$ are rather limited. 

The analytical analysis based on angular- or $\lambda$-selections is intended to prove the SBR-performance of the imager using experimental data. However, the efficiency reduction  implied by this type of analysis obviously represents an important drawback for the practical implementation of the proposed method. For this reason, an alternative analysis technique has been developed, which is discussed in the second part of this article. 

In summary, a significant reduction of the background level is obtained in the 10~keV neutron energy range for the measurement of the $^{56}$Fe($n,\gamma$) reaction with i-TED, when compared to state-of-the-art \cds detectors. An additional background suppression effect is achieved after implementing selections in the $\gamma$-ray imaging domain, although the counting statistics are at the limit to extract accurate information. The results obtained here are limited, to a large extent, due to the limitations in hardware and reconstruction software used in these tests. We expect to improve these aspects both by means of additional A-modules as shown in Ref.~\cite{Babiano20}, and by means of improved position-reconstruction techniques~\cite{Balibrea20}. The expected performance of the final i-TED detector after these recent upgrades is discussed in the following Sec.~\ref{sec:Prospects}. In addition, this Section explores alternative analysis algorithms, which help to mitigate significantly the limitation in detection efficiency related to the imaging cuts.

%
%
\section{Prospects for background suppression: the final \fpis and innovative methods}\label{sec:Prospects}
\subsection{Perspectives on background suppression with \fpis}\label{sec:ImagingiTED5}
The previous sections have presented the first experimental proof-of-concept on background suppression using an i-TED prototype with limited hardware characteristics and position-reconstruction algorithms compared to the final \fpis detector, hereafter simply referred to as i-TED. The latter will be composed of four individual Compton Imaging modules, each of the them consisting on two position sensitive detection layers based on monolithic LaCl$_{3}$(Ce) crystals. Each scintillator is optically coupled to a SiPM, which features
a segmentation of 8x8 pixels. The S plane is composed of a single 50x50x10 mm$^{3}$ crystal, while the A plane consists of an array of four crystals, each of them with a size of 50x50x25 mm$^{3}$. A detailed description of the first demonstrator of an i-TED module can be found in Ref.~\cite{Babiano20}.

  To illustrate the impact of the aforementioned upgrades, the detection efficiency and background suppression perspectives with the final i-TED detector have been studied on the basis of MC simulations described in Sec.~\ref{sec:MCiTED5}. In Sec.~\ref{sec:iTED4piImaging} we present the expected performance of i-TED in terms of background suppression.

\subsubsection{MC simulations of i-TED }\label{sec:MCiTED5}

The final i-TED is currently under construction and characterization~\cite{Babiano20} and the position-reconstruction algorithms have recently undergone a major upgrade based on the use of Machine Learning techniques~\cite{Balibrea20}. In this context, Monte Carlo (MC) simulations become an essential tool for testing the sensitivity of the detector performance to the improved features, developing new background suppression methods and planning the upcoming neutron capture time-of-flight experiments~\cite{Se79Proposal}. Initial MC simulations of the background suppression concept were presented in Ref.~\cite{Domingo16}. To evaluate the prospects and capabilities of the final detector, more accurate MC simulations including the details in the geometry as well as experimental effects have been carried out in this work. The simulations have been made with the \textsc{Geant4}~\cite{Geant4_2} toolkit (v10.6) using the officially released QGSP\_BIC\_HP Physics List~\cite{Geant4PL} which contains the standard Electromagnetic Package.

\begin{figure}[!htpb]
   \begin{center}     \includegraphics[width=0.8\columnwidth]{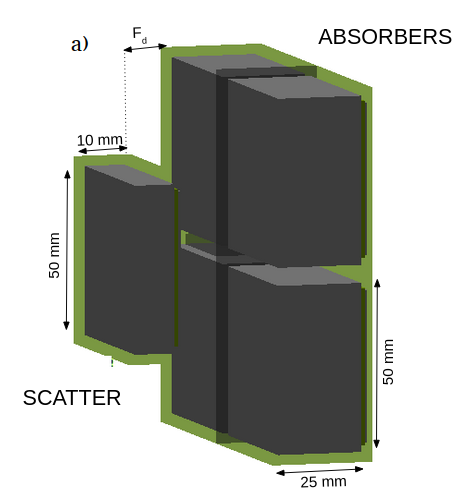}
   \includegraphics[width=0.6\columnwidth]{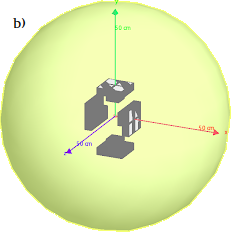}
      \end{center}
\caption{\label{fig:MCiTED} Schematic view of each of the four i-TED modules as implemented in \textsc{Geant4} indicating the dimensions of the LaCl$_{3}$(Ce) crystals of the Scatter and Absorber layers (a). General view of the i-TED detector surrounded by the sphere from which the background \g-rays are emitted (b).}
\end{figure}

The general view of each of the i-TED modules as implemented in \textsc{Geant4} is displayed in Fig.~\ref{fig:MCiTED}~(a). A critical parameter of this system is the distance between the two detection planes, so-called focal distance, which was set to 15~mm for all the simulations in this work. According to a previous work~\cite{Babiano20}, this value ensures a good balance between efficiency an angular resolution. Each of the four i-TED modules was placed at a different distance (25, 50, 75 and 100~mm) from the sample to study its impact in the signal-to-background ratio (SBR).

A realistic study of the expected background suppression using i-TED requires an accurate modelling of both the capture cascades and the background events. The well known $^{197}$Au(n,\g) cascade was chosen to simulate capture events. The \g-rays emitted in the de-excitation of $^{198}$Au were generated using the Captugens code~\cite{Captugen}, which combines the known level scheme with an statistical description of the compound nucleus, as explained in Ref.~\cite{Guerrero2008}. The $^{197}$Au(n,\g) cascades were emitted within the \textsc{Geant4} application from a 0.1~mm thick Au disc with a diameter of 20~mm placed at the expected sample position in the center of i-TED (see Fig.~\ref{fig:MCiTED}) to replicate the actual experimental conditions at n\_TOF. As for the simulation of the extrinsic neutron-induced background, the input for the \textsc{Geant4} simulations was the experimental \g-ray spectrum registered with the absorber crystal of the i-TED prototype described in Sec.~\ref{sec:iTED} during a measurement of a carbon sample, a pure neutron scatter. These \g-rays, coming in the real experiment from neutrons scattered in the sample and captured in the walls of the experimental area, were emitted from a random position on a sphere of 1~m in diameter surrounding i-TED (see Fig.~\ref{fig:MCiTED}~(b)).

 A total of 10$^{7}$ $^{197}$Au(n,\g) capture events and 3$\times$10$^{9}$ background events were simulated to register about 10$^{5}$ coincidence events of each type in the individual detectors. The output of the MC simulation resembles that of the experimental set-up, including for each simulated event the deposited energy, interaction position and time of all the \g-ray hits in the S- and A-layers of the four i-TED modules. Experimental effects such as the low energy threshold, position and energy resolutions were included in the simulations to consider their impact in the imaging resolution (see Ref.~\cite{Babiano20}) and consequently in the performance of the background suppression method. 

\subsubsection{Compton imaging method for background suppression with i-TED: the analytical approach}\label{sec:iTED4piImaging}

 The background reduction perspectives with i-TED with respect to the state-of-the-art \cds detectors have been studied on the basis of MC simulations introduced in the previous section. For the C$_{6}$D$_{6}$, the same (n,\g) and background events were simulated using the \textsc{Geant4} application described in Ref.~\cite{Lerendegui16} and the  C$_{6}$D$_{6}$ set-up geometry of previous experiments at n\_TOF-EAR1~\cite{Lerendegui18}. 
  
       \begin{figure}[!htb]
   \begin{center}     \includegraphics[width=\columnwidth]{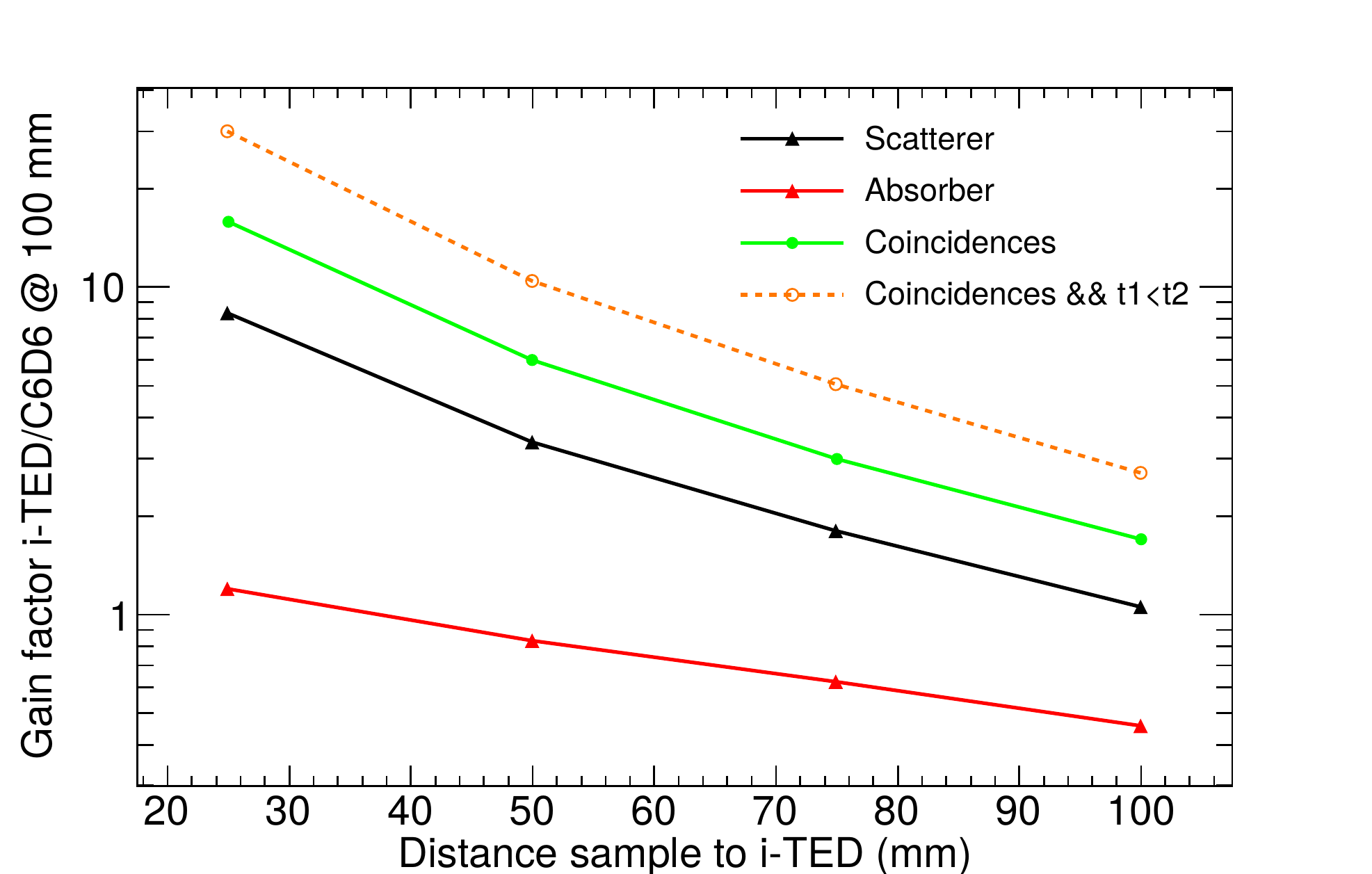}
      \end{center}
\caption{\label{fig:GainCoincidencesC6D6} Expected SBR gain factor with respect to a \cds detector at 10~cm as a function of the i-TED distance to the sample. The scatter (black triangles) and absorber (red triangles) crystals alone are compared to one i-TED module operated in coincidence mode (green solid) and after selection of events with t$_{1}$ $< $t$_{2}$ (orange dashed) (see text for details).}
\end{figure}

 \begin{figure}[!htbp]
 \begin{center}     \includegraphics[width=\columnwidth]{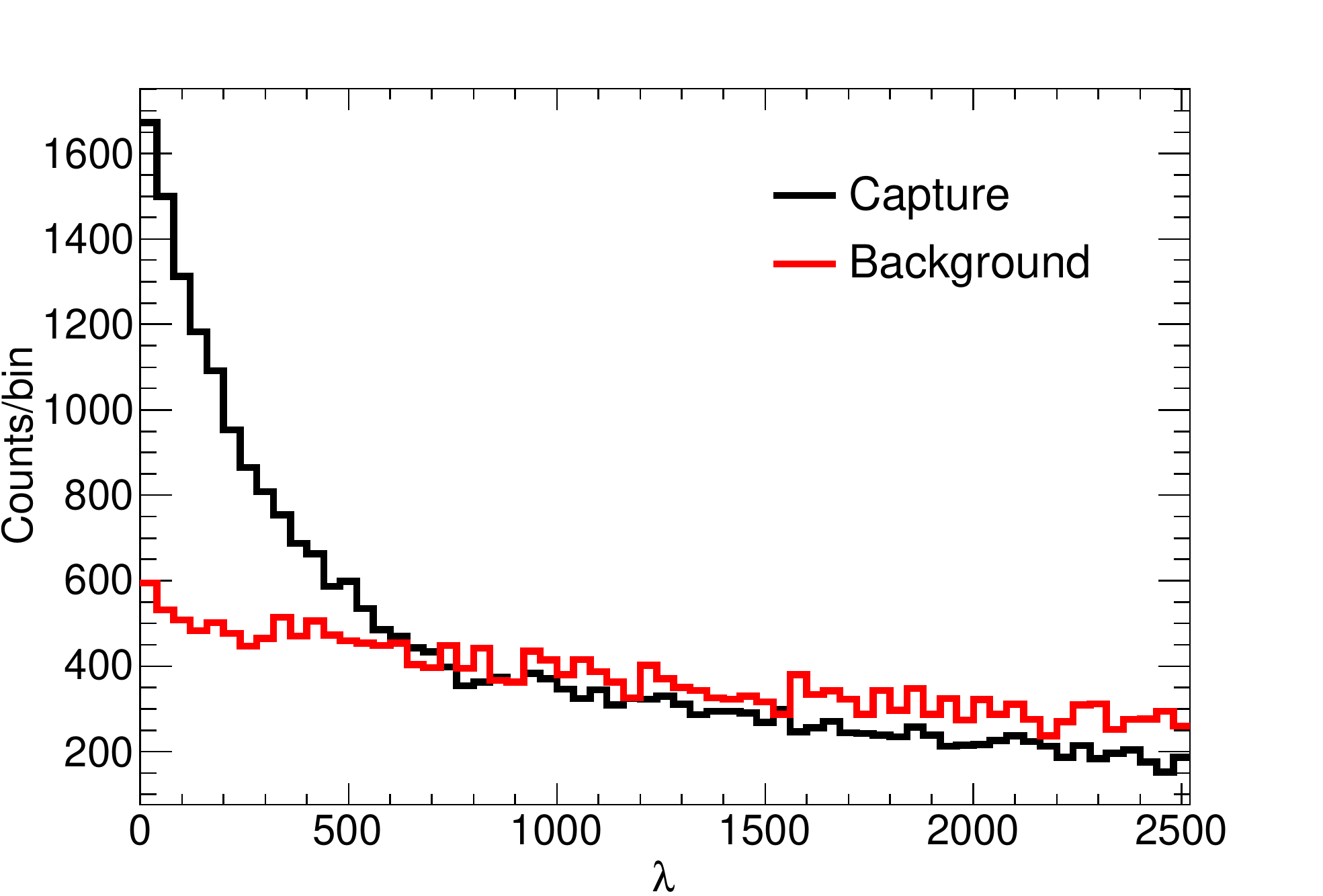}
      \end{center}
\caption{\label{fig:LambdaCaptBckg} Distribution of the imaging parameter $\lambda$ for capture and background events for one module of i-TED placed at 50~mm from the sample. Both distributions have been normalized to the integral over the full $\lambda$ range.}
\end{figure}

The results of these MC study indicate that the upgrades in the final i-TED detector will enhance not only the efficiency, as it was mentioned in Sec.~\ref{sec:TOFiTED}, but also the background reduction effect with respect to the commissioned prototype.

As it was mentioned in Sec.~\ref{sec:iTEDsensitivity}, the background suppression capability of i-TED is not only based on the application of selections in the $\gamma$-ray imaging domain. The operation of the S- and A-planes in time-coincidence, required to apply the Compton technique, also enhances the SBR. The MC simulations of the final i-TED, confirm the effect of the time-coincidence mode in the SBR, as it is shown in Fig.~\ref{fig:GainCoincidencesC6D6}. This figure displays the expected SBR of an i-TED module at different distances in single- and coincidence-mode relative to a C$_{6}$D$_{6}$ detector at 10~cm. For \cds and i-TED at the same distance, one can appreciate that the scatter detector in singles-mode would present a similar SBR than a \cd, while the operation in coincidence is expected to improve the SBR in a 60\%. According to Fig.~\ref{fig:GainCoincidencesC6D6}, for an i-TED module placed at 63~mm from the sample, replicating the setup of the commissioned prototype, the time-coincidence mode will yield a 4 times higher SBR than a \cd, improving the results obtained with the i-TED demonstrator (see Fig.~\ref{fig:EnFeiTEDvsC6D6imaging}).

The results in Fig.~\ref{fig:GainCoincidencesC6D6} indicate also that selecting the events in which the scatter (t$_{1}$) is fired before the absorber (t$_{2}$) improves the SBR in an additional factor 1.8, due to the different spatial origin of (n,\g) and background events. The relevance of the time discrimination has triggered an on-going study to optimize the coincidence resolving time (CRT) of i-TED, which will be presented in future works.

\begin{figure}[!htbp]
   \begin{center}     \includegraphics[width=\columnwidth]{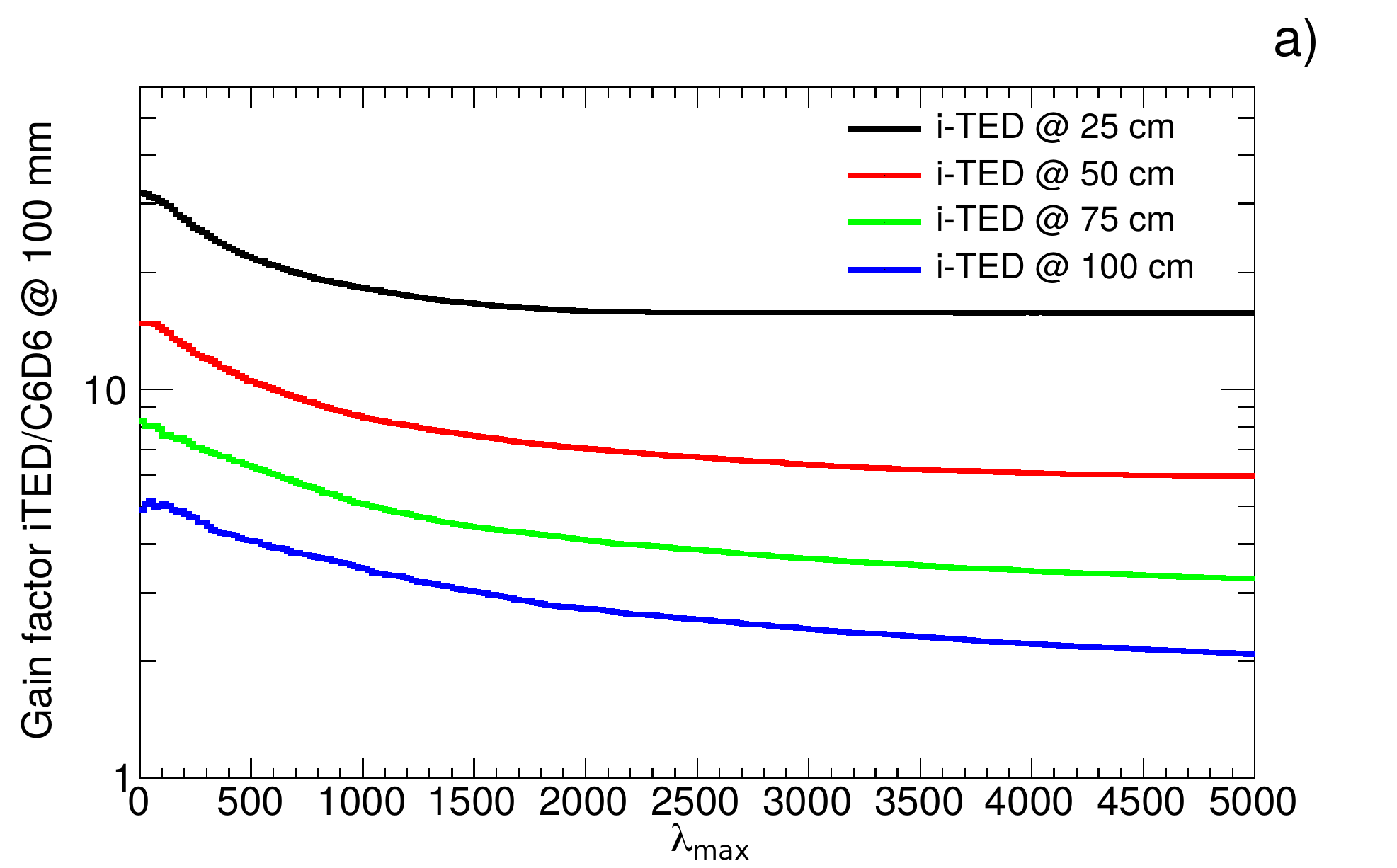}
   \includegraphics[width=\columnwidth]{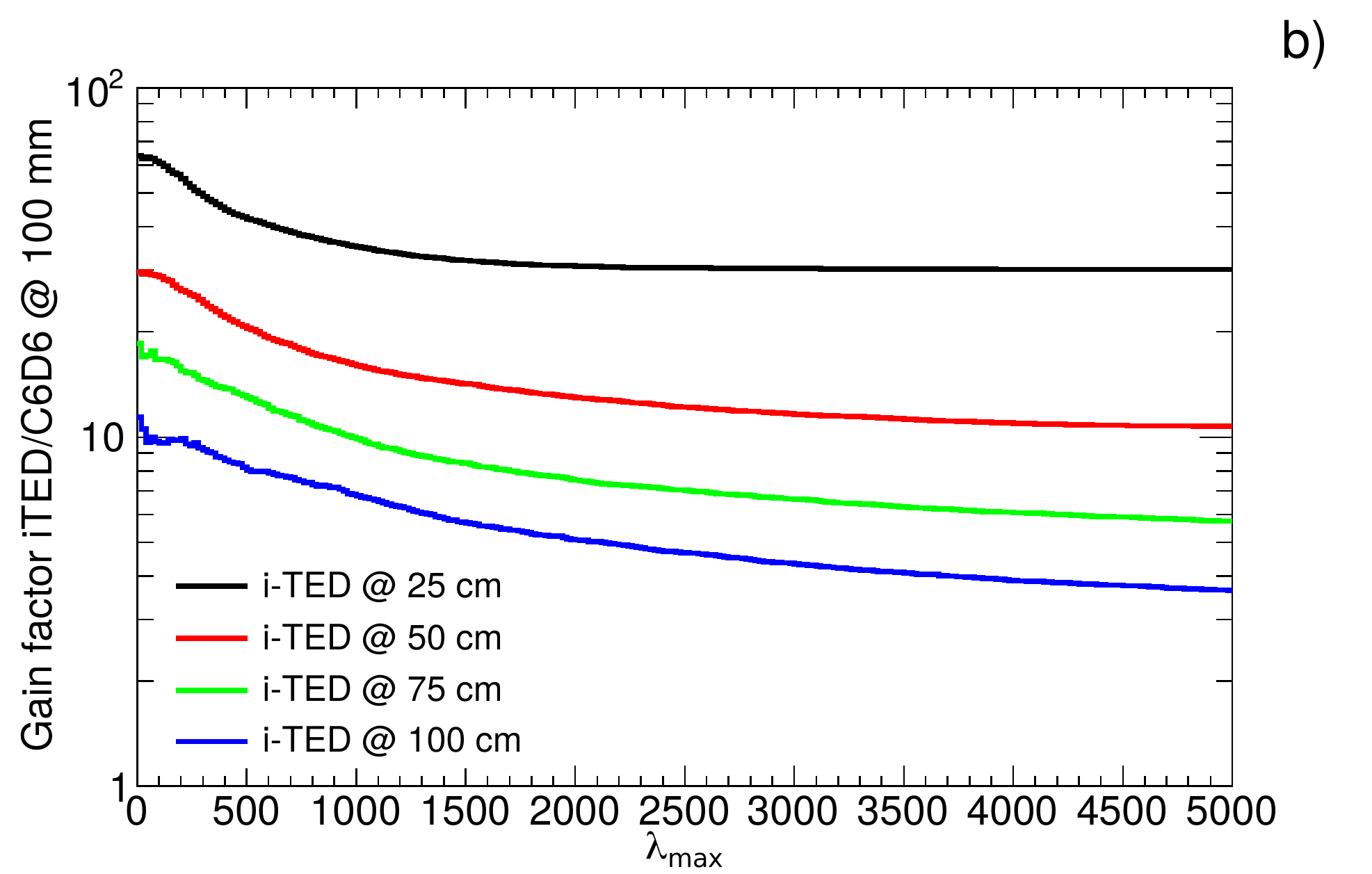}
      \end{center}
\caption{\label{fig:GainImagingC6D6} Total SBR gain factor with respect to a \cds detector at 10 cm as a function of the $\lambda$ cut. Each curve corresponds to a different distances between the i-TED module and the sample. The two panels correspond to the expected performance of i-TED with no CRT (a) and with an ideal CRT that allows selecting events with t1$<$t2 (b).}
\end{figure}
 
The application of Compton Imaging techniques for background suppression, introduced in Sec.~\ref{sec:iTEDsensitivity}, is based on setting selections in the imaging parameter $\lambda$, defined in Eq.~(\ref{eq:Lambda}). The expected $\lambda$  distribution for capture and background events extracted from the MC simulations of i-TED is displayed in Fig.~\ref{fig:LambdaCaptBckg}. From this figure it is clear that selecting events below a given lambda ($\lambda_{max}$=500-1000) leads to an enhanced SBR.

By combining the coincidence-mode and the imaging selections, the total expected SBR gain of i-TED with respect to a C$_{6}$D$_{6}$ detector is presented in Fig.~\ref{fig:GainImagingC6D6} for modules placed at different distances. In this figure, panel a) presents a conservative scenario with time information (no CRT), while in panel b) we show the best SBR after selecting only events with t1$<$t2 with an ideal CRT. The experimental CRT, still under optimization, will yield an intermediate situation. 

Our MC study indicates that i-TED with an imaging cut of e.g. $\lambda_{max}$=500 would yield an SBR gain factor between 4 and 10 (depending on the achievable CRT and the choice of $\lambda_{max}$) with respect to a C$_{6}$D$_{6}$ placed at the same distance. The background can be further suppressed by placing the i-TED modules closer to the sample. Indeed, for the sample to detector distance used in the i-TED PoC measurements (63~mm), a background ratio of 8 (no CRT) to 15 (ideal CRT) would be expected. These values indicate that the experimentally proven SBR gain ($<=3.5$) using the commissioned prototype of only three crystals (see Fig.~\ref{fig:EfficiencySBratioFeiTED}) would be significantly improved with the final detector of 20 crystals. The large impact of the detector positioning in the background suppression shown in Fig.~\ref{fig:GainImagingC6D6} will be experimentally investigated during the commissioning phase of i-TED at CERN n\_TOF in 2021.
 
The major drawback of the imaging-based background suppression is the drop of efficiency, discussed in Fig.~\ref{fig:EfficiencySBratioFeiTED}, due to the lost (n,$\gamma$) events after applying a given $\lambda_{max}$ cut. This could represent a major limitation for establishing this technique in TOF facilities, where beam-time needs to be shared over several experiments every year. Therefore, a radically new approach based on Machine Learning (ML) algorithms has been explored and tested on the basis of the MC simulation of the final i-TED. This is presented in the following section.

\subsection{ML algorithms for background rejection with i-TED}\label{sec:MLBackground}

Exploiting the background suppression capability of i-TED while preserving a larger fraction of the capture efficiency is the aim of testing innovative methods beyond the analytical imaging cuts presented before. In particular, the high time- and energy-resolution and the granularity of i-TED provide an ideal test-bench for ML techniques, when compared to other detectors commonly used in (n,\g) experiments. From the physics perspective, capture gamma-ray events in i-TED yield a pattern, which is spatially inward and energetically hard. This signature is, on average, quite different from the backward and softer spectrum related to background gamma-rays, that originate after neutron capture in the surroundings of the set-up. This makes i-TED especially suited to benefit from the power of ML techniques, that can be trained to differentiate between one pattern and another.

A brief introduction to the methodology followed in this work to test ML algorithms for background suppression is presented in Sec.~\ref{sec:MLIntro}. Sec.~\ref{sec:MLResults} presents the perspectives for background rejection with i-TED using ML discrimination. Last, preliminary tests of ML-based background suppression on experimental data are presented in Sec.~\ref{sec:MLExpResults}.

\subsubsection{Machine Learning: algorithms and methodology}\label{sec:MLIntro}

The background suppression with i-TED can be simplified as a problem of binary classification in the framework of ML techniques. In this work, the performance of several state-of-the-art ML algorithms included in \textit{Scikit-learn} Python module~\cite{scikit-learn} has been evaluated: k-Nearest neighbors, Logistic Regression, Support Vector Classifier (SVC), Gaussian Naive Bayes (NB), Random Forest, XGBoost Classifier and Keras.

To train the ML algorithms in the discrimination of capture and background events, the same number (50k~-~100k) of (n,\g) and background events were selected from the MC output individually for each i-TED module. Each MC event (S\&A coincidence) contains the same nine features determined with the detector in a real measurement: 3D coordinates of the \g-ray interactions in the two PSDs (6), energy deposited in the S- and A-planes (2) and time difference between both interactions in scatter and absorber (1). Additionally, the Compton angle, calculated from the deposited energies, and the $\lambda$ parameter, calculated using Eq.~(\ref{eq:Lambda}) from these parameters, were also included in the training to improve the classification performance. 

 \begin{figure}[!htbp]
   \begin{center}     \includegraphics[width=0.7\columnwidth]{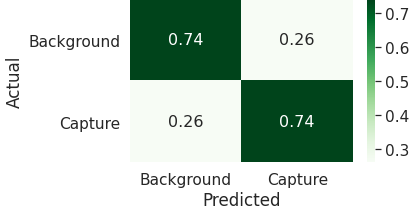}
      \end{center}
\caption{\label{fig:ConfusionMatrix} Confusion matrix showing the prediction accuracy of the XGBoost classifier to discriminate capture and background events (see text for details). This result corresponds to the best scenario with ideal CRT.}
\end{figure}

The MC-based input data-set was split, being 75\% of the events devoted to training the ML classifier and the remaining 25\% to testing its accuracy. The performance of a ML classifier is typically expressed in the so-called Confusion Matrix~\cite{ConfusionMatrix}, which represents the fraction of events of either type (capture and background in our case) that are correctly and wrongly classified by the ML algorithm. Fig.~\ref{fig:ConfusionMatrix} shows an example of Confusion Matrix obtained in this work for the best performing algorithm, XGBoost (eXtreme Gradient Boosting)~\cite{XGBoost} classifier. An optimization of the algorithm based on the minimization of the \textit{log loss} function lead to the following parameters:
\begin{itemize}
    \item The number of gradient boosted trees in the model was set to \texttt{n\_estimators} = 150. 
    \item  A \texttt{learning\_rate} = 0.1. This parameter shrinks the weight of each feature after each boosting step to prevent over-fitting and is optimized in parallel to the number of trees.
    \item A tree depth of \texttt{Max\_depth} = 8 determines the complexity of the model or the ability to learn relations very specific to a particular data sample.
\end{itemize}
The remaining parameters were proven to have a negligible impact in the outcome of the classifier. Similar results are obtained using deep neural networks with the Keras deep learning framework~\cite{Keras}, while other algorithms featured worse performances (see Table~\ref{TableML}).

\begin{table}[htb!]
\begin{center}
\caption{Accuracy of the different ML classifiers tested in this work, defined as the total fraction of correctly classified background and capture events. The results correspond to the best scenario of MC events with ideal CRT.}
\begin{tabular}{lc} 
\hline \hline
\textbf{ML algorithm}               &  \textbf{Best accuracy (\%)}\\
\hline
k-Nearest neighbors                 &   69.2                    \\
Support Vector Classifier (SVC)     &   69.5                           \\
Gaussian Naive Bayes (NB)           &   69.9                           \\
Logistic Regression                 &   70.1                           \\
Random Forest                       &   72.7                           \\
XGBoost Classifier                  &   74.3                     \\
Keras                               &   74.3                           \\
\hline \hline    
\end{tabular} 
\label{TableML}
\end{center}
\end{table}

\subsubsection{ML-based background suppression with i-TED}\label{sec:MLResults}

The ML algorithm showing the best performance, the XGBoost classifier, has been applied to study the perspectives of background suppression with i-TED. This algorithm was trained to discriminate capture and background events from the MC simulation following the methodology described in the previous section. In a similar way than in the imaging-based method (see Fig.~\ref{fig:GainImagingC6D6}), two scenarios in terms of time resolution have been considered: MC events with ideal CRT and MC events with no time information.

Two Figures of Merit have been extracted from the Confusion Matrix (see Fig.~\ref{fig:ConfusionMatrix}) and used to evaluate the performance of the ML classifier:
\begin{itemize}
    \item \textbf{(n,\g) efficiency}: fraction of capture events which are correctly classified (bottom right entry).
    \item \textbf{SBR gain factor}: fraction of correct (n,\g) event (bottom right) over the fraction of background events wrongly predicted as capture (top right).
\end{itemize}

\begin{figure}[!htbp]
   \begin{center}     \includegraphics[width=\columnwidth]{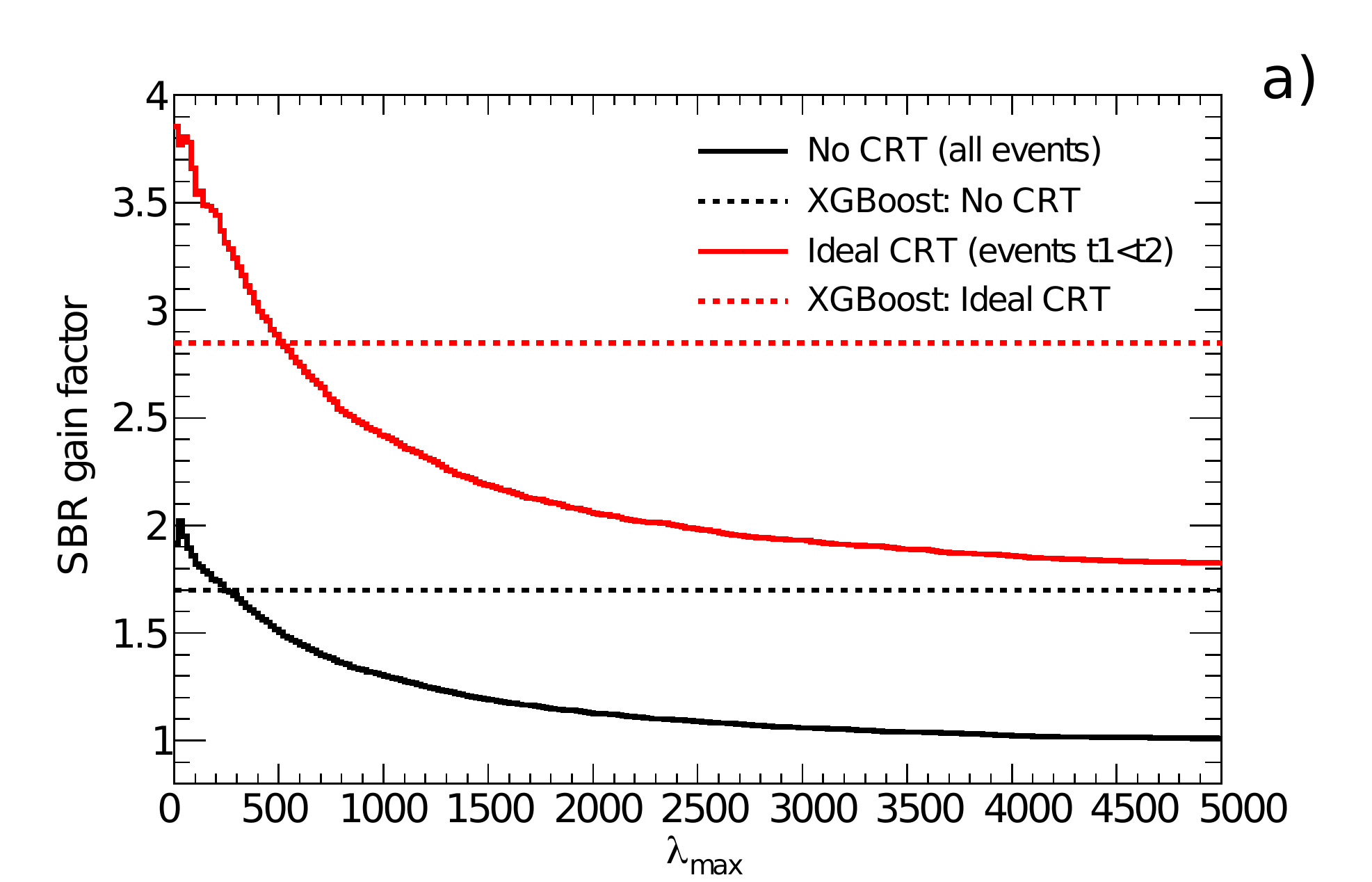}
   \includegraphics[width=\columnwidth]{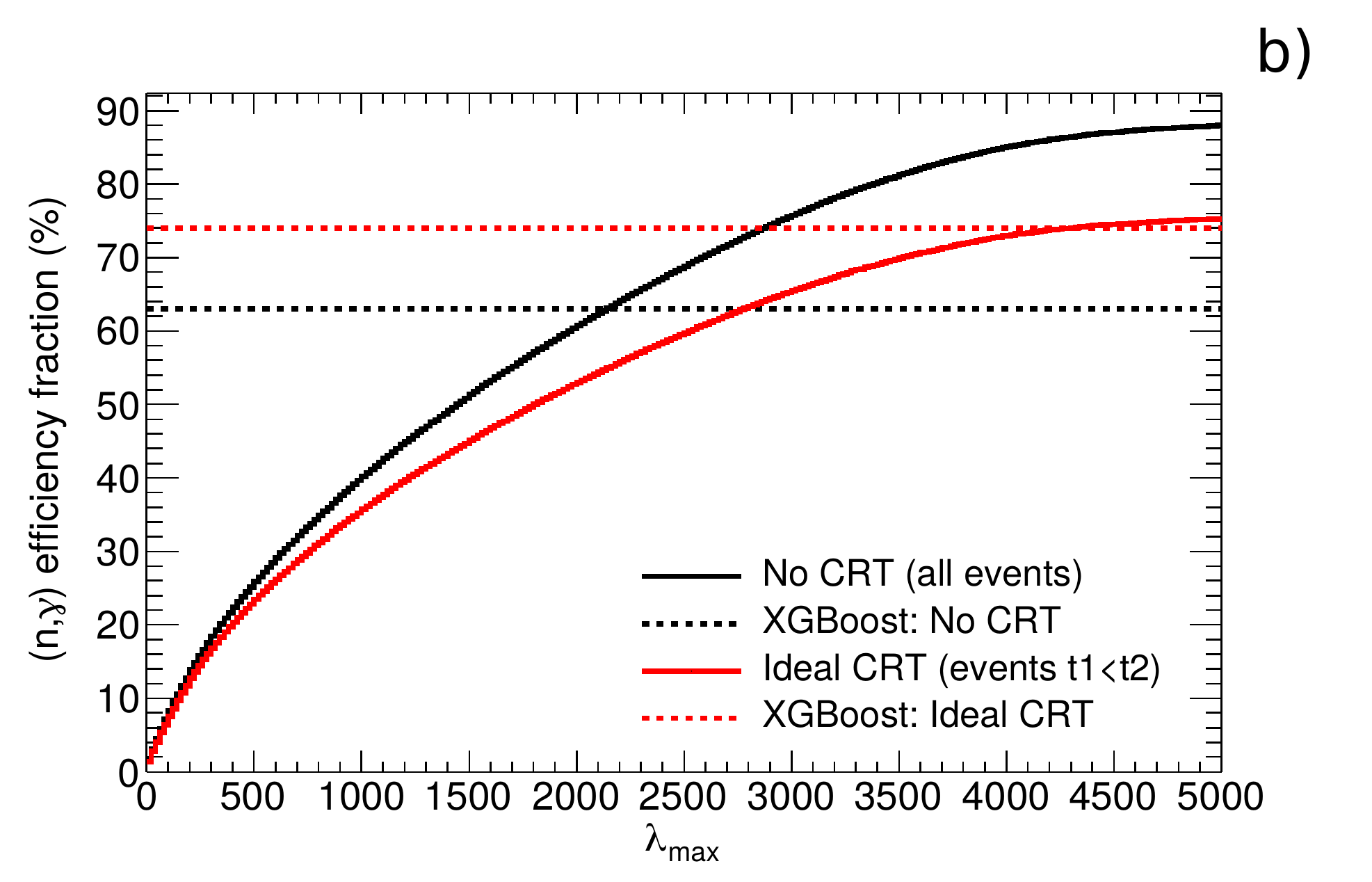}
      \end{center}
\caption{\label{fig:MLvsImaging} SBR gain factor (a) and fraction of the efficiency for capture events as a function of the $\lambda$ imaging cut (b). The black and red solid lines correspond to the analytical imaging case with no time resolution and ideal CRT, respectively. The dashed lines show the results obtained with XGBoost.}
\end{figure}

The performance of the XGBoost classifier is compared to the imaging-based results in Fig.~\ref{fig:MLvsImaging} for the two CRT scenarios. The top panel shows the additional SBR gain factor related to the imaging or ML selection on top of the background reduction related to the operation of i-TED in coincidence-mode (see Fig.~\ref{fig:GainCoincidencesC6D6}). The level of background reduction obtained with XGBoost is equivalent to an imaging selection with $\lambda_{max}$ that ranges between 300 and 600 depending on the time resolution scenario. From Fig.~\ref{fig:MLvsImaging} one concludes that an ideal CRT also enhances substantially the performance of the ML-based background suppression from a SBR of 1.7 to 2.8. An intermediate result can be expected for the experimental scenario and it will depend on the CRT experimentally attainable and the separation between the two detection planes.

The main advantage of the ML-based method compared to the imaging approach is demonstrated in panel (b) of Fig.~\ref{fig:MLvsImaging}. The ML-based technique keeps 63 to 74\% of the efficiency (depending on the CRT), a factor between 2 and 3 higher than the imaging method with the same SBR gain factor. 

According to the results presented in this section, Machine Learning methods seem very promising for background suppression with i-TED, providing a similar SBR and a clear efficiency enhancement with respect to the imaging method. However, the final application of ML methods for background suppression in (n,\g) experiments should be based on experimental data rather than MC events since the MC modelling of the detector response may not include all the experimental features. Preliminary tests of ML-based background suppression based on experimental data taken with the prototype described in Sec.~\ref{sec:iTED} are presented in the following section.

\subsubsection{ML-based background suppression on i-TED prototype data}\label{sec:MLExpResults}

 The same XGBoost classifier described in Sec.~\ref{sec:MLIntro} has been tested using experimental data measured during the commissioning of the aforedescribed i-TED prototype at n\_TOF-EAR1. In a first step, our aim is to demonstrate experimentally the higher (n,\g) efficiency and the similar background suppression of ML- and imaging-based methods.
 
The training of the ML algorithm was carried out using two separate measurements. The same number (8~kEvents) of capture and background events were extracted from measurements of $^{197}$Au and $^{nat}$Pb samples, respectively. For the case of $^{197}$Au, events were selected from the 4.9~eV saturated resonance, where the background represents less than 0.1\% of the events. On the other hand, neutron capture in Pb is negligible and this sample can be considered a pure neutron and \g-ray scatter, both contributing to the overall background at n\_TOF~\cite{Zugec14}.

\begin{figure}[!htbp]
   \begin{center}     
  \includegraphics[width=\columnwidth]{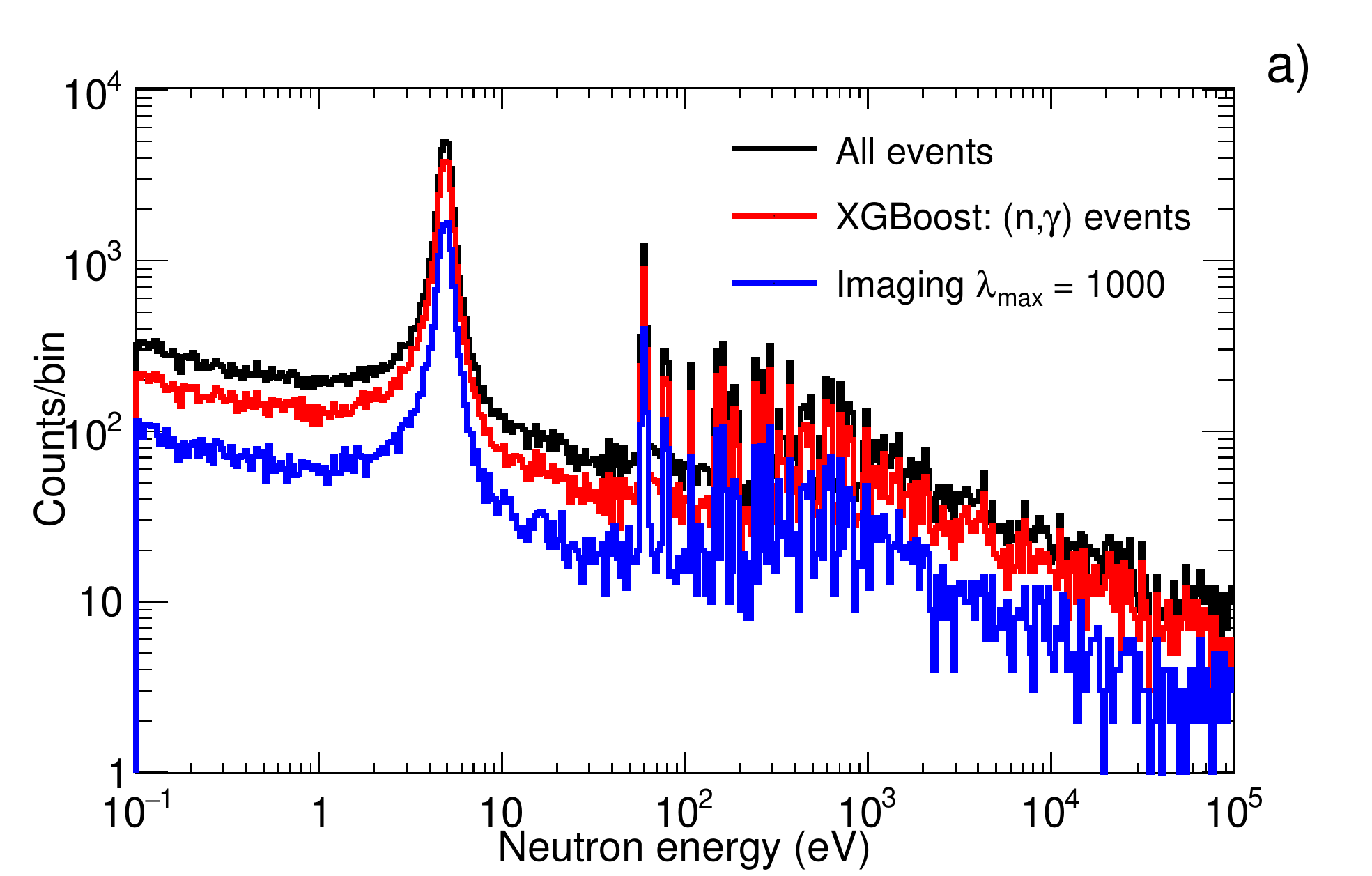}
   \includegraphics[width=\columnwidth]{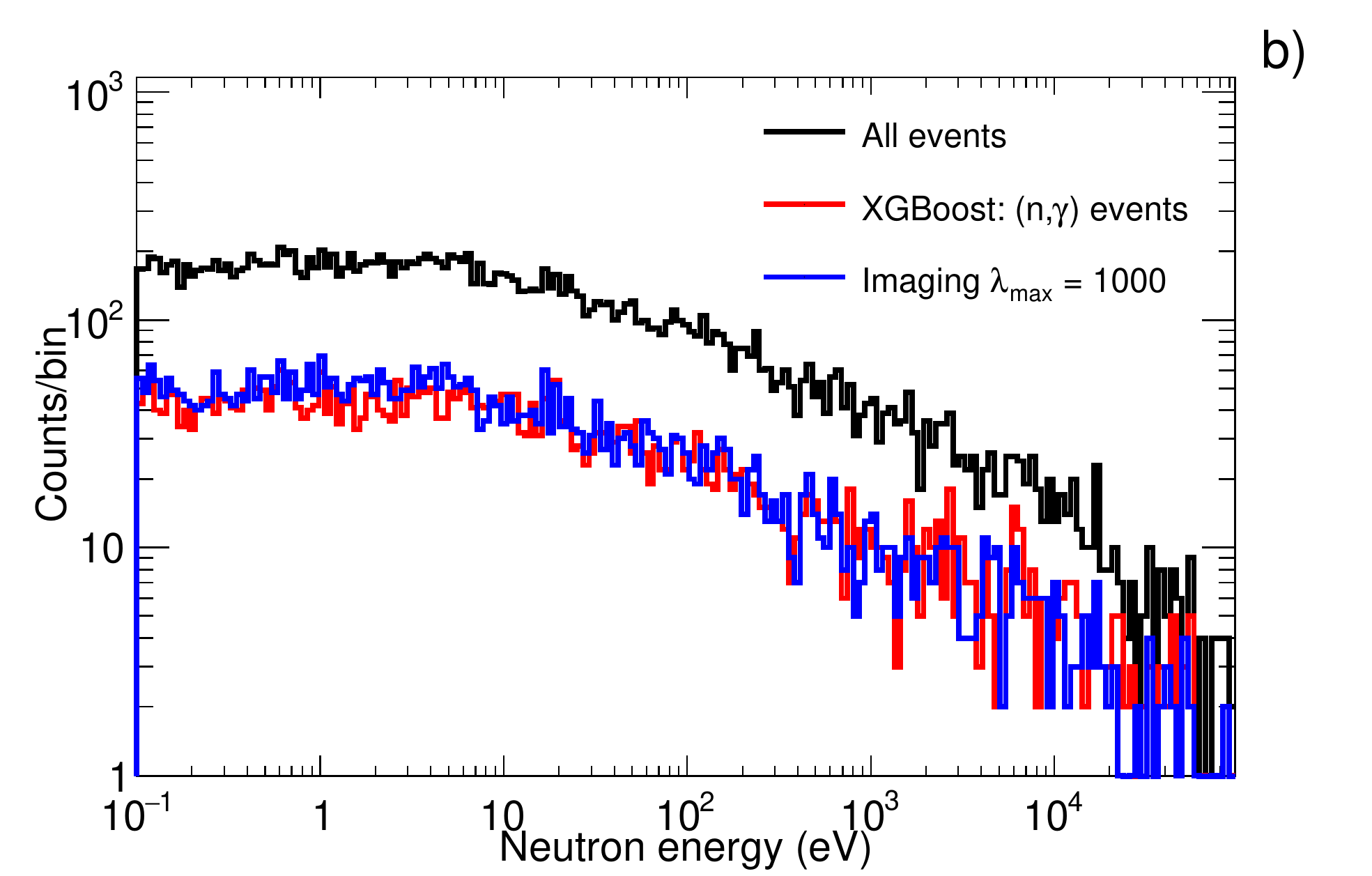}
      \end{center}
\caption{\label{fig:MLExpData} Counting rate as a function of the neutron energy measured with $^{197}$Au (a) and a C-nat (b) samples. The original data set (black) is compared to the fraction of events identified as capture by the XGBoost algorithm (red) and the events selected with an imaging cut $\lambda_{max}$=1000.}
\end{figure}

To prove the higher (n,\g) efficiency of the ML-classifier with respect to the standard imaging method, we applied the trained algorithm to the full $^{197}$Au(n,\g) data set and selected only those events predicted as being capture. The result is shown in top panel of Fig.~\ref{fig:MLExpData}, where the resulting counting rate of $^{197}$Au(n,\g) obtained with the XGBoost classifier is compared to the original data set and the results with an imaging cut $\lambda_{max}$=1000. From this figure one concludes that the efficiency for $^{197}$Au(n,\g) is almost 75-80\% in the whole energy range, clearly improved with respect to the 30\% obtained with the selection in the imaging domain.

The same XGBoost classifier has been applied to a background-like measurement obtained with a carbon sample which, similarly to the case of Pb, is a pure neutron scatter. The events classified as capture by the ML algorithm, presented in the bottom panel of Fig.~\ref{fig:MLExpData}, show a reduction in 65\% with respect to the original background level, very similar to the reduction obtained by selecting events with $\lambda_{max}$=1000.

In summary, the first tests of ML-based background suppression on experimental data confirm the higher (n,\g) efficiency and similar reduction of the background with respect to the experimentally proven imaging method. At this stage, the only limitation to explore further the performance of the ML method is the limited statistics available in the data-set discussed in Sec.~\ref{sec:iTEDsensitivity}. With only $\sim$2000 counts in the 1.15~keV $^{56}$Fe+n resonance (in coincidence mode) it does not become feasible to train the algorithm reliably. This fact prevented us from applying the ML-method to the $^{56}$Fe data set and compare the SBR gain with the ML approach and the results from the analytical method shown in Sec.\ref{sec:iTEDsensitivity}. The further development and experimental validation of the ML methods will be completed in the upcoming commissioning of the full i-TED at CERN n\_TOF in 2021.

\section{Summary and outlook}\label{sec:summary}

i-TED is an innovative total energy detector which exploits Compton imaging techniques to enhance the signal-to-background ratio (SBR) in (n,\g) time-of-flight experiments. This paper has discussed the status and perspectives of the background rejection capabilities using i-TED. 

The first experimental validation of the background reduction in a $^{56}$Fe(n,\g) measurement at CERN n\_TOF using a previous prototype has been presented. This result is based on a comparison with state-of-the-art C$_{6}$D$_{6}$ detectors to benchmark the performance of this novel methodology and apparatus. Despite of the use of a prototype under development, a SBR gain of a factor 2.5-3.7 with respect to a C$_{6}$D$_{6}$ has been experimentally demonstrated in a neutron capture measurement on $^{56}$Fe, confirming the applicability of this concept. Moreover, this work has shown that the performance of the scatter crystals in singles-mode yields an efficiency and SBR performance comparable to that of a state-of-the-art C$_{6}$D$_{6}$ detector, with the additional strength of the better energy resolution in \lacl.

In light of the experimentally validated SBR gain, this work has explored the background rejection prospects for the final i-TED design. Accurate MC simulations of the final detector have shown that a two-fold increase in efficiency and a factor 3-5 improvement in SBR with respect to the commissioned prototype is expected. This MC study has also highlighted the large impact of the CRT in the reduction of the neutron-induced background with this system. 

The main drawback of the analytical imaging method, the sharp drop in (n,\g) efficiency, is clearly improved using new methods based on Machine Learning (ML) techniques, which are very powerful thanks to the complexity of this Compton Imager. In this work we have shown that these ML algorithms (XGBoost) applied to MC data keep the (n,$\gamma$) efficiency loss below 30\% compared to the 70\% experimentally reported using Compton imaging. Last, preliminary tests of the ML-based background rejection using experimental data from the prototype commissioning confirm  a similar performance than analytical imaging cuts in terms of background reduction, while the efficiency for capture events is about 3 times larger than with the analytical approach. 

The last steps in the development of the final i-TED detector and the implementation of the enhanced position reconstruction techniques will be carried out in the forthcoming year. Moreover, neutron sensitivity studies at HiSPANoS-CNA are also planned. Last, the commissioning of the detector at CERN n\_TOF will provide the experimental validation of the new ML-methodology, prior to the first neutron cross section experiments with i-TED, planned for 2022. 

\section*{Declaration of competing interest}
The authors declare that they have no known competing financial interests or personal relationships that could have appeared to influence the work reported in this paper.

\section*{CRediT authorship contribution statement}
\textbf{V. Babiano-Su\'arez:} Investigation, Methodology, Formal analysis, Data curation, Visualization,  Writing - original draft.
\textbf{J. Lerendegui-Marco:} Investigation, Methodology, Conceptualization, Formal analysis, Data curation, Visualization, Writing - original draft. 
\textbf{J. Balibrea-Correa:} Investigation, Methodology, Formal analysis, Data curation.
\textbf{D. Calvo:} Investigation. 
\textbf{L. Caballero:} Investigation, Methodology. 
\textbf{I. Ladarescu:} Software, Visualization.
\textbf{C. Domingo-Pardo:} Conceptualization, Methodology, Supervision, Writing -review \& editing, Funding acquisition, Investigation, Formal analysis.
\textbf{F. Calvi{\~n}o:} Investigation.
\textbf{A.~Casanovas:} Investigation.
\textbf{A.~Tarife\~no-Saldivia:} Investigation.
\textbf{V. Alcayne:} Investigation.
\textbf{C.~Guerrero:} Investigation.
\textbf{M.A. Mill\'an-Callado:} Investigation.
\textbf{M.T. Rodr\'iguez-Gonz\'alez:} Investigation.
\textbf{M.~Barbagallo:} Investigation.
\textbf{Other co-authors n\_TOF Collaboration:} Investigation, Resources,  Writing review.

\section*{Acknowledgment}
This work has been carried out in the framework of a project funded by the European Research Council (ERC) under the European Union's Horizon 2020 research and innovation programme (ERC Consolidator Grant project HYMNS, with grant agreement n$^{\circ}$ 681740). The authors acknowledge support from the Spanish Ministerio de Ciencia e Innovaci\'on under grants PID2019-104714GB-C21, FPA2017-83946-C2-1-P, FIS2015-71688-ERC and CSIC for funding PIE-201750I26. We would like to thank the crew at the Electronics Laboratory of IFIC, in particular Manuel Lopez Redondo and Jorge N\'acher Ar\'andiga for their excellent and efficient work.


\end{document}